\documentclass[aps,twocolumn,superscriptaddress,preprintnumbers,nofootinbib]{revtex4-1}
\pdfoutput=1

\usepackage{pslatex}
\usepackage[pdftex]{graphicx}
\usepackage{psfrag}
\usepackage{epsfig}
\usepackage{color}
\usepackage{cancel}
\usepackage{slashed}
\usepackage{amssymb}
\usepackage{amsmath}
\usepackage{hyperref}
\usepackage{enumerate}
\usepackage{multirow}
\usepackage{natbib} 
\usepackage{comment} 

\begin{document}

\title{Implications of first neutrino-induced nuclear
  recoil measurements in direct detection experiments}%
\author{D. Aristizabal Sierra}%
\email{daristizabal@uliege.be}%
\affiliation{Universidad T\'ecnica Federico Santa Mar\'{i}a -
  Departamento de F\'{i}sica\\ Casilla 110-V, Avda. Espa\~na 1680,
  Valpara\'{i}so, Chile}%
\author{N. Mishra}%
\email{nityasa_mishra@tamu.edu}%
\affiliation{Department of Physics and Astronomy, Mitchell Institute
  for Fundamental Physics and Astronomy, Texas A\&M University,
  College Station, Texas 77843, USA}%
\author{L. Strigari}%
\email{strigari@tamu.edu}%
\affiliation{Department of Physics and Astronomy, Mitchell Institute
  for Fundamental Physics and Astronomy, Texas A\&M University,
  College Station, Texas 77843, USA}%
\begin{abstract}
  PandaX-4T and XENONnT have recently reported the first 
  measurement of nuclear recoils induced by the $^8$B solar neutrino
  flux, through the coherent elastic neutrino-nucleus scattering (CE$\nu$NS) channel. 
  As long anticipated, this is an important milestone
  for dark matter searches as well as for neutrino physics. This measurement means 
  that these detectors have reached exposures such that searches for low mass, 
  $\lesssim 10$ GeV dark matter cannot be analyzed using the background-free paradigm
  going forward. It also opens a new era for these detectors to
  be used as neutrino observatories. In this paper we assess the
  sensitivity of these new measurements to new physics in the neutrino
  sector. We focus on neutrino non-standard interactions
  (NSI) and show that---despite the still moderately low statistical
  significance of the signals---these data already provide valuable
  information. We find that limits on NSI from PandaX-4T and XENONnT
  measurements are comparable to those derived using combined
  COHERENT CsI and LAr data, as well as those including the latest Ge
  measurement. Furthermore, they provide sensitivity to pure $\tau$ flavor
  parameters that are not accessible using stopped-pion or reactor
  sources. With further improvements of statistical uncertainties as
  well as larger exposures, forthcoming data from these experiments
  will provide important, novel results for CE$\nu$NS-related physics.
\end{abstract}
\maketitle

\section{Introduction}
\label{sec:intro}
PandaX-4T~\cite{PandaX:2024muv} and XENONnT~\cite{Aprile:2024rhz} have
recently reported the detection of coherent elastic neutrino-nucleus
scattering (CE$\nu$NS) induced by $^8$B solar neutrinos. Due to their
low energy thresholds and large active volumes these experiments
identify the $^8$B component of the solar neutrino flux at a
significance level of the order of $2\sigma$. This is the first
detection of CE$\nu$NS from an astrophysical source, complementing the
recent detections from the stopped-pion source by the COHERENT
experiment~\cite{COHERENT:2017ipa,COHERENT:2019iyj,Adamski:2024yqt}. Further,
this detection probes the CE$\nu$NS cross section at characteristic
neutrino energy scales lower than that probed by COHERENT and with a
new material target\footnote{Measurements at COHERENT have employed
  CsI, LAr and more recently Ge. Both PandaX-4T and XENONnT, instead,
  rely on LXe.}.

The detection of solar neutrinos at dark matter (DM) detectors such as
PandaX-4T and XENONnT is a milestone in neutrino
physics~\cite{Monroe:2007xp,Vergados:2008jp,Strigari:2009bq,Billard:2013qya,OHare:2021utq}. It
represents an important step in the continuing development of the
solar neutrino program, dating back to over half of a century. From
the perspective of solar neutrino physics, it is the second pure
neutral current channel detection of the solar neutrino flux,
complementing the SNO neutral current detection of the flux using a
deuterium target~\cite{SNO:2011hxd}. Its observation was anticipated
long time ago to be not only a challenge for DM searches, but also an
opportunity for a better understanding of neutrino properties and
searches of new physics \cite{Aalbers:2022dzr}.

The detection of $^8$B neutrinos via CE$\nu$NS has important
implications more broadly for neutrino physics, astrophysics, and DM.
This detection has the potential to provide information on the
properties of the solar interior~\cite{Cerdeno:2017xxl}. It also has
the potential to probe new physics in the form of non-standard
neutrino interactions
(NSI)~\cite{Dutta:2017nht,AristizabalSierra:2017joc,AristizabalSierra:2019ykk,Dutta:2019oaj},
sterile neutrinos~\cite{Billard:2014yka,Alonso-Gonzalez:2023tgm},
neutrino electromagnetic properties
\cite{AristizabalSierra:2020zod,AristizabalSierra:2021fuc,Cadeddu:2020nbr,Giunti:2023yha}
or new interactions involving light mediators
\cite{AristizabalSierra:2019ykk,AristizabalSierra:2020edu}. Detection
of solar neutrinos via CE$\nu$NS also is important for interpreting
the possible detection of low mass, $\lesssim 10$ GeV, dark
matter~\cite{Dent:2016wor,AristizabalSierra:2021kht}. A detailed
understanding of this signal is of paramount importance for the
interpretation of future data. The identification of a possible WIMP
signal requires a thorough understanding of neutrino-induced nuclear
recoils.

In this paper, we examine the sensitivity of the PandaX-4T and XENONnT
data to NSI. We show that even with this early data these measurements
are already capable of providing competitive bounds. In
particular, because of neutrino flavor conversion, these measurements
are sensitive to all neutrino flavors and so open flavor channels not
accessible in CE$\nu$NS experiments relying on $\pi^+$ decay-at-rest or
reactor neutrino fluxes. Thus from this point of view these
experiments are very unique.

The reminder of this paper is organized as follows. In
Sec. \ref{sec:new_physics} we discuss the Standard Model (SM) CE$\nu$NS cross section,
define the parameters we use in our calculation and briefly discuss
the experimental input employed. In Sec. \ref{sec:new_physics} we
provide a detailed discussion of NSI effects in both propagation and
detection. To do so we rely on the two-flavor approximation, which
provides rather reliable results up to corrections of $\sim 10\%$
\footnote{Note that both PandaX-4T and XENONnT data have statistical
  uncertainties of the order of $37\%$
  \cite{PandaX:2024muv,Aprile:2024rhz}. Theoretical precision below
  $10\%$ is therefore at this stage not required.}. In
Sec. \ref{sec:analysis}, after briefly discussing the main features of
both PandaX-4T and XENONnT data, we present the results of our
analysis. Finally, in Sec. \ref{sec:conclusions} we summarize and
present our conclusions. In App.~\ref{sec:limits_summary} we provide a
summary of NSI limits arising from the one-parameter analysis.
\section{CE$\nu$NS cross section, $^8$B solar neutrino flux, event
  rates and experimental input}
\label{sec:new_physics}
In the SM, at tree-level the CE$\nu$NS cross section
is has no lepton flavor dependence~\cite{Freedman:1973yd}, with
flavor-dependent corrections appearing at the one-loop
level~\cite{Sehgal:1985iu,Tomalak:2020zfh,Mishra:2023jlq}. At tree
level the scattering cross section reads \cite{Freedman:1973yd}
\begin{equation}
  \label{eq:cross-sec-NSI}
  \frac{d\sigma}{dE_r}=  \frac{G_F^2}{2\pi}\,Q^2_W\,m_N
  \left(2 - \frac{m_NE_r}{E_\nu^2}\right)F^2_W(E_r)\ .
\end{equation}
For the nuclear mass, $m_N$, we use the averaged mass number
$\langle A\rangle=\sum_{i=1}^9X_iA_i$, where $i$ runs over the nine
stable xenon isotopes and $X_i$ refers to $i$-th isotope natural
abundance.  $Q_W$ refers to the weak charge and determines the
strength at which the $Z$ gauge boson couples to the nucleus. At
tree-level and neglecting $q^2$ dependent terms ($q$ referring to the
transferred momentum) the weak charge is entirely determined by the vector neutron and
proton couplings
\begin{equation}
  \label{eq:Q_W}
  Q_W= Z g^p_V + N\,g^n_V\ ,
\end{equation}
with $Z=54$ referring to the nucleus atomic number and
$N=(\langle A\rangle - Z)$ to the number of neutrons. The nucleon
couplings are in turn given by the fundamental electroweak neutral
current up and down couplings: $g_V^p=1/2 - 2\sin^2\theta_W$ and
$g_V^n=-1/2$. Because of the value of the weak mixing
angle\footnote{In our analysis we use the SM central value prediction
  extrapolated to low energies ($q=0$),
  $\sin^2\theta_W=0.23857\pm 0.00003$ \cite{Kumar:2013yoa}.}, $g_V^n$
exceeds $g_V^p$ by more than a factor 20. Thus, up to small
corrections the total cross section scales as $N^2=(A-Z)^2$.

Effects due to the finite size of the nucleus are parameterized in
terms of the weak-charge form factor, $F_W$, for which different
parametrizations can be adopted. However, given the energy scale of
solar neutrinos these finite size nuclear effects are small, not
exceeding more than a few percent regardless of the parametrization
\cite{AristizabalSierra:2019zmy,Sierra:2023pnf}. Although of little
impact, our calculation does include the weak-charge form factor.  We
have adopted the Helm parametrization \cite{Helm:1956zz} along with
$R_n=R_C + 0.2\,\text{fm}$, with $R_C$ calculated by averaging the
charge radius of each of the nine xenon stables isotopes over their
natural abundance.

\begin{table}
  \renewcommand{\arraystretch}{1.5}
  \setlength{\tabcolsep}{8pt}
  \centering
  \begin{tabular}{|c||c|c|}\hline
    Flux & Normalization [$\text{cm}^{-2}\text{s}^{-1}$] & End-point [MeV]\\\hline\hline
    $pp$ & $5.98\times 10^{10}$&$0.40$\\\hline
    $^7\text{Be}$ &$4.93\times 10^9$&$0.38,0.86$\\\hline
    $pep$ &$1.44\times 10^8$&$1.44$\\\hline
    $^{13}\text{N}$&$2.78\times 10^8$&$1.20$\\\hline
    $^{15}\text{O}$&$2.05\times 10^8$&$1.73$\\\hline
    $^{17}\text{F}$&$5.29\times 10^6$&$1.74$\\\hline
    $^8\text{B}$&$5.46\times 10^6$&$16.0$\\\hline
    $hep$&$7.98\times 10^3$&$18.7$\\\hline
  \end{tabular}
  \caption{Neutrino flux normalization as recommended in
    Ref. \cite{Baxter:2021pqo} and inline with the B16(GS98) SSM. For
    detection only $^8$B matters. For propagation we include the whole
    spectrum.}
  \label{tab:normalization}
\end{table}
$^8$B electron neutrinos are produced in $\beta^+$ decay processes:
$^8\text{B}\to \text{Be}^*+e^++\nu_e$. The features of the spectrum as
well as its normalization is dictated by the Standard Solar Model
(SSM). In our analysis we use the values predicted by the B16(GS98)
SSM \cite{Vinyoles:2016djt}. The distribution of $^8$B neutrino
production from the B16(GS98) SSM peaks at around
$5\times 10^{-2}\,R_\odot$ and ceases to be efficient at
$0.1\,R_\odot$, where the distribution fades away. For the calculation
of event rates only the $^8$B neutrino flux is required. For the
calculation of propagation effects (matter effects), however, we
require all possible fluxes. In all cases we adopt neutrino spectra
normalization as recommended for reporting results for direct DM
searches \cite{Baxter:2021pqo}, which are inline with those predicted
by the B16(GS98) SSM. The values for those normalization factors along
with the kinematic end-point energies for all fluxes are shown in
Tab.~\ref{tab:normalization}.

Calculation of differential event rate spectra follows from
convoluting the CE$\nu$NS differential cross section in
Eq. (\ref{eq:cross-sec-NSI}) with the $^8$B spectral function, namely
\begin{equation}
  \label{eq:DRS}
  \frac{dR}{dE_r}=\frac{\varepsilon\,N_A}{m_\text{mol}^\text{Xe}}
  N_{^8\text{B}}\int_{E_\nu^\text{min}}^{E^\text{max}_\nu}
  \frac{d\Phi_{^8\text{B}}}{dE_\nu}
  \frac{d\sigma}{dE_r}dE_\nu\ .
\end{equation}
Here $\varepsilon$ refers to exposure measured in tonne-year, $N_A$ is
the Avogadro number in 1/mol units,
$m_\text{mol}^\text{Xe}=131.3\times 10^{-3}\text{kg/mol}$,
$N_{^8\text{B}}$ the $^8$B flux normalization from Tab.
\ref{tab:normalization}, $E^\text{min}_\nu=\sqrt{m_N E_\nu/2}$ and
$E^\text{max}_\nu$ the kinematic end-point of the $^8$B spectrum from
Tab.~\ref{tab:normalization} as well. Eq.~(\ref{eq:DRS}) is valid in
the SM, where the CE$\nu$NS differential cross section is flavor
universal at tree level. If either through one-loop corrections or new
physics the cross section becomes flavor dependent, then the integrand
should involve the probability associated with each neutrino flavor
(see Sec. \ref{sec:detection_effects} for a more detailed
discussion). The event rate follows from integration of
Eq. (\ref{eq:DRS}) over recoil energies, with the experimental
acceptance $\mathcal{A}(E_r)$ fixed according to the PandaX-4T or
XENONnT data sets. Generically it reads
\begin{equation}
  \label{eq:event_rate}
  R = \int_{E_r^\text{min}}^{E_r^\text{max}}\mathcal{A}(E_r)
  \frac{dR}{dE_r}dE_r\ .
\end{equation}

PandaX-4T perform two types of analyses on their data. First, they
perform a combined S1/S2 analysis, in which a neutrino signal event is
identified via both prompt scintillation and secondary ionization
signals from the nuclear recoil (\textit{paired signal}). The low
energy threshold for this analysis is set by the S1 signal, which in
terms of nuclear recoil energy is $\sim 1.1$ keV. The second analysis
is an S2 only analysis, in which only the ionization component is used
as the signal of an event (\textit{US2 signal}). In this case, the
nuclear recoil threshold is lower, $\sim 0.3$ keV, but the trade-off
is an increase in the backgrounds for this sample.

PandaX-4T present data from two runs: their commissioning run, which
they call Run0, and their first science run, which they call Run1. For
the paired data set, the exposure is 1.25 tonne-year, and for the US2,
the exposure is 1.04 tonne-year. Using a maximum likelihood analysis,
PandaX-4T finds a best fitting $^8$B event rate from the US2 sample of
$75\pm 28$ and a paired event rate of $3.5 \pm 1.3$.

The XENONnT collaboration combined two separate analyses, labelled SR0
and SR1, which when combined amount to an exposure of 3.51
tonne-year. They present acceptances for both an S1 only and an S2
only analysis. For the primary analysis, XENONnT combine the
acceptances for S1 and S2 (with a resulting 0.5 keV threshold), and,
for this combined exposure, they quote a best fit event rate of
$10.7_{-4.2}^{+3.7}$. They point out that this result is in close
agreement with: (i) Expectations from the measured solar $^8$B
neutrino flux from SNO, (ii) the theoretical CE$\nu$NS cross section
with xenon nuclei, (iii) calibrated detector response to low-energy
nuclear recoils. For the expected event rate, they find 
$11.9^{+4.5}_{-4.2}$. Calculation of the $Z$-score---assuming these
results to be independent---yields $0.2\,\sigma$. Thus using either in
our statistical procedure produces no sizable deviation in the final
results. Tab. \ref{tab:detectors_summary} summarizes the detector
parameter configurations along with the signals we have employed.
\begin{table}
  \renewcommand{\arraystretch}{1.5}
  \centering
  \begin{tabular}{|c||c|c|c|}\hline
    Data set & Exp [tonne-year] & $E_r^\text{min,max}$ [keV] & Signal \\\hline\hline
    PandaX-4T (paired) & 1.25 & 1.1/3.0 & $3.5 \pm 1.3$\\\hline
    PandaX-4T (US2) & 1.04 & 0.3/3.0 & $75\pm 28$\\\hline
    XENONnT & 3.51 & 0.5/3.0 & $10.7\pm 3.95$\\\hline
  \end{tabular}
  \caption{PandaX-4T (paired and US2) and XENONnT parameter detector
    configurations used in the NSI statistical analysis. Values taken
    from Refs. \cite{PandaX:2024muv,Aprile:2024rhz}.}
  \label{tab:detectors_summary}
\end{table}
\section{Neutrino non-standard interactions} 
\label{sec:NSI}
In addition to loop-level corrections, flavor-dependence in the
CE$\nu$NS cross section may also be introduced through neutrino
NSI~\cite{Barranco:2005yy}. The effective Lagrangian accounting for
the new vector interactions can be written as
\begin{equation}
  \label{eq:NSI_Lag}
  \mathcal{L}_\text{NSI}=-\sqrt{2}G_F\sum_{\substack{i=e,\mu,\tau\\q=u,d}}
  \overline{\nu}_i\,\gamma_\mu P_L
  \epsilon_{ij}^q\,\nu_j\,\overline{q}\gamma^\mu q\ ,
\end{equation}
where the $\epsilon_{ij}^q$ parameters determine the strength of the
effective interaction with respect to the SM strength. Neutrino NSI
affect neutrino production, propagation and detection. Since
production takes place through charged-current (CC) processes, effects
in production are small \footnote{For instance, off-diagonal CC NSI
  can induce charged lepton rare decays for which stringent bounds
  apply. Diagonal CC NSI can induce contact $e^{-}e^{+}q\overline{q}$
  interactions for which collider limits apply too.}. Effects on
propagation and detection, being due to neutral current, can instead
be potentially large. Thus we consider only those two. Propagation
effects arise from forward scattering processes which induce matter
potentials proportional to the number density of the scatterers. So in
addition to the SM matter potential, the new interaction---being of
vector type---induces additional matter potentials that affect
neutrino propagation and thus neutrino flavor conversion. Detection,
instead, becomes affected because of the impact of the new effective
interaction on the CE$\nu$NS cross section. All in all, NSI effects on
solar neutrinos may be prominent in propagation $\oplus$ detection.

Neutrino NSI are constrained by a variety of experimental
searches. Here we provide a summary of the main constraints, which
does not aim at being complete but rather to provide a general picture
of what has been done (for a more detailed account see
e.g. Ref. \cite{Farzan:2017xzy}). First of all, global analysis of
oscillation data imply tight constraints on the size and flavor
structure of matter effects. Thus, those constraints can be translated
into limits on NSI parameters
\cite{Gonzalez-Garcia:2011vlg,Gonzalez-Garcia:2013usa}. Limits
involving global analysis of oscillation data combined with CE$\nu$NS
measurements have been also derived
\cite{Esteban:2018ppq,Coloma:2023ixt}. Constraints from CE$\nu$NS data
alone, for which only effects on detection apply, have been analyzed
using both CsI data releases along with LAr data in
Ref. \cite{DeRomeri:2022twg}, and also the most recent
measurement with germanium in Ref. \cite{Liao:2024qoe}. Further
constraints from monojets and missing energy searches at the LHC
exist~\cite{Friedland:2011za,BuarqueFranzosi:2015qil}. Involving
electrons and at early times, the new interaction can keep neutrinos
in thermal contact with electrons and positrons below $\sim
1\,$MeV. Requiring small departures from this value leads to
cosmological constraints \cite{deSalas:2021aeh}. In supernov\ae,
neutrino NSI have as well been considered in
e.g. Refs. \cite{Esteban-Pretel:2007zkv,Jana:2024lfm}.

In what follows we describe their effects in propagation and in
detection. To do so we rely on the two-flavor approximation, well
justified up to corrections of the order of $10\%$ because of
$\Delta m_{12}^2/\Delta m_{13}^2\ll 1$ and $\sin^2\theta_{13}\ll 1$
\cite{deSalas:2020pgw}. And rather than including the data and
constraints discussed above, we focus only on the constraints implied
by PandaX-4T and XENONnT.
\subsection{Neutrino NSI: Propagation effects}
\label{sec:propagation_effects}
Electron neutrinos are subject to flavor conversion in the Sun,
governed by the vacuum and matter Hamiltonians
\begin{equation}
  \label{eq:evolution}
  i\frac{d}{dr}|\boldsymbol{\nu}\rangle
  =\left[
    \frac{1}{2E_\nu}\boldsymbol{U}\;
    \boldsymbol{H_\text{vac}}\;\boldsymbol{U}^\dagger
    +
    \boldsymbol{H_\text{mat}}
  \right]|\boldsymbol{\nu}\rangle\ .
\end{equation}
Here $|\boldsymbol{\nu}\rangle^T= |\nu_e, \nu_\mu, \nu_\tau \rangle^T$
refers to the neutrino flavor eigenstate basis, $r$ to the neutrino
propagation path,
$\boldsymbol{U}=U_{23}U_{13}U_{12}\equiv U(\theta_{23})U(\theta_{13})U(\theta_{12})$
is the $3\times 3$ leptonic mixing matrix parametrized in the standard
way,
$\boldsymbol{H_\text{vac}}=\text{diag}(0,\Delta m_{21}^2,\Delta
m_{31}^2)$ and in the absence of NSI the matter Hamiltonian is given
by
$\boldsymbol{H_\text{mat}}=\sqrt{2}G_F\,n_e(r)\text{diag}(1,0,0)$. Note
that because of matter potentials neutrino flavor evolution is more
conveniently followed in the flavor basis.

As previously pointed out, the presence of neutrino NSI induce new
matter potential terms that modify the flavor evolution equation,
namely
\begin{equation}
  \label{eq:ev-eq}
  i\frac{d}{dr}|\boldsymbol{\nu} \rangle
  =
  \left[
    \frac{1}{2E_\nu}\boldsymbol{U}\;
    \boldsymbol{H_\text{vac}}\;
    \boldsymbol{U}^\dagger
    +
    \sqrt{2}G_Fn_e(r)
    \sum_{f=e,u,d}
    \boldsymbol{\varepsilon^f}
  \right]|\boldsymbol{\nu} \rangle\ ,
\end{equation}
where the NSI coupling matrices $\varepsilon^f$ involves the quark
relative abundances in addition to the parameters entering in
Eq. (\ref{eq:NSI_Lag}):
\begin{equation}
  \label{eq:varepsilon-matrix}
   \boldsymbol{\varepsilon^f}=
   \begin{pmatrix}
     1 + \varepsilon_{ee}^f & \varepsilon_{e\mu}^f & \varepsilon_{e\tau}^f \\
     \varepsilon_{e\mu}^f & \epsilon_{\mu\mu}^f & \varepsilon_{\mu\tau}^f\\
     \varepsilon_{e\tau}^f & \varepsilon_{\mu\tau}^f  & \varepsilon_{\tau\tau}^f\\
   \end{pmatrix}\ .
\end{equation}
Explicitly, $\varepsilon_{ij}^f(r)=Y_f(r)\epsilon_{ij}^f$ ($f=e,u,d$)
with $Y_f(r)=n_f(r)/n_e(r)$. The up- and down-quark relative
abundances are written in terms of the neutron relative abundance
$Y_u = 2 + Y_n$ and $Y_d = 1 + 2 Y_n$, with the neutron number density
calculated from the $^4$He and $^1$H mass fractions.

A three-flavor analysis of NSI matter effects demands numerical
integration of Eq. (\ref{eq:ev-eq}) for each point in the NSI
parameter space. However, an analytical, less CPU expensive and yet
precise approach can be adopted in the so-called mass dominance limit
$\Delta m_{13}^2\to\infty$ \cite{Gonzalez-Garcia:2013usa}. In this
approximation, neutrino propagation is properly described in the basis
$|\widetilde{\nu}\rangle = \mathcal{U}^T|\nu\rangle \equiv
U_{13}^TU_{23}^T|\nu\rangle$ (\textit{propagation basis}). Up to
corrections of the order of $\sin\theta_{13}$, the propagating
neutrino states are: A mainly electron neutrino state ($\tilde\nu_e$),
a superposition of muon and tau neutrinos state ($\tilde\nu_\mu$) and
its orthogonal counterpart ($\tilde\nu_\tau$). With these
considerations, only $\tilde\nu_e$ and $\tilde\nu_\mu$ have sizable
mixing. Mixing with $\tilde\nu_\tau$ for neutrino energies of the
order of $10\,$MeV and average SSM quark number densities does not
exceed $3\times 10^{-2}\times \epsilon_{ij}^q$
\cite{AristizabalSierra:2017joc}. With $\tilde\nu_\tau$ ``decoupled''
from mixing, flavor conversion becomes then a two-flavor problem that
can be entirely treated at the analytic level.

In two-flavor approximation, the survival probability is given by
$\mathcal{P}_{ee}(E_\nu,r)$~\cite{Gonzalez-Garcia:2013usa}
\begin{equation}
  \label{eq:survival-prob}
  \mathcal{P}_{ee}(E_\nu,r)=\cos^4\theta_{13}
  \,\mathcal{P}_\text{eff}(E_\nu,r)
  +
  \sin^4\theta_{13}\ ,
\end{equation}
where the $r$ dependence is introduced by the effective probability
given by \cite{Parke:1986jy}
\begin{equation}
  \label{eq:eff-prob}
  \mathcal{P}_\text{eff}(E_\nu,r)=
  \frac{1+\cos2\theta_M(r)\cos2\theta_{12}}{2}\ .
\end{equation}
Here $\theta_M(r)$ is the mixing angle in matter and adiabatic
propagation has been assumed, thus implying a rather suppressed
level-crossing probability ($P_c\to 0$). With neutrino oscillation
data taken from Ref.~\cite{deSalas:2020pgw}, calculation of the survival
probability in Eq. (\ref{eq:survival-prob}) then reduces to the
determination of $\theta_M$. To do so the following $2\times 2$
Hamiltonian has to be diagonalized
\begin{equation}
  \label{eq:2times2-ham}
  \boldsymbol{H}=\frac{1}{4E_\nu}
  \begin{pmatrix}
    -\Delta m_{21}^2 \cos2\theta_{12} + A 
    & \Delta m_{21}^2 \sin2\theta_{12} + B\\
    \Delta m_{21}^2 \sin2\theta_{12} + B 
    & \Delta m_{21}^2 \cos2\theta_{12} - A
  \end{pmatrix}\ .
\end{equation}
In this expression the $A$ and $B$ terms in the diagonal and
off-diagonal entries are given by
\begin{align}
  \label{eq:A-B}
  A&=4\sqrt{2}E_\nu G_F n_e(r) 
     \left[\frac{\cos^2\theta_{13}}{2} - Y_q(r)\varepsilon_D^q\right]\ ,
     \nonumber\\
  B&=4\sqrt{2}E_\nu G_F n_e(r) Y_q(r)\varepsilon_N^q \ ,
\end{align}
from where it can be seen that in the limit $\epsilon_{ij}^q=0$ and
$\cos\theta_{13}=0$, $A$ reduces to the SM term and $B$ vanishes. The
parameters $\varepsilon_D$ and $\varepsilon_N$ result from the
rotation from the flavor to the propagation basis and read
\cite{Gonzalez-Garcia:2013usa}:
\begin{widetext} 
\begin{align}
  \label{eq:epsilonD}
  \varepsilon_D^q&=
  -\frac{c_{13}^2}{2}\epsilon_{ee}^q
  +\frac{\left[c_{13}^2 - \left(s_{23}^2 - s_{13}^2 c_{23}^2\right)\right]}{2}
  \epsilon_{\mu\mu}^q
  +\frac{\left(s_{23}^2 - c_{23}^2s_{13}^2\right)}{2}
  \epsilon_{\tau\tau}^q + s_{13}c_{13}s_{23}\epsilon_{e\mu}^q
  + s_{13}c_{13}c_{23}\epsilon_{e\tau}^q
  - (1+s^2_{13})\,c_{23}s_{23}\epsilon_{\mu\tau}^q\ ,
  \\
  \label{eq:epsilonN}
  \varepsilon_N^q&= 
  -s_{13}c_{23}s_{23}\epsilon_{\mu\mu}^q
  + s_{13}c_{23}s_{23}\epsilon_{\tau\tau}^q
  + c_{13}c_{23}\epsilon_{e\mu}^q
  - c_{13}s_{23}\epsilon_{e\tau}^q
  + s_{13}\left(s_{23}^2 - c_{23}^2\right)\epsilon_{\mu\tau}^q\ ,
\end{align}
\end{widetext} 
where $c_{ij}\equiv \cos\theta_{ij}$ and
$s_{ij}\equiv \sin\theta_{ij}$. The mixing angle in matter thus can be
written as
\begin{equation}
  \label{eq:mixing-angle}
  \cos2\theta_M(r) = \frac{\Delta m_{12}^2\cos2\theta_{12}-A}
  {\sqrt{\left(\Delta m_{12}^2\cos2\theta_{12}-A\right)^2
      +
  \left(\Delta m_{12}^2\sin2\theta_{12}+B\right)^2}}\ .
\end{equation}
\begin{figure}[h]
  \centering
  \includegraphics[scale=0.35]{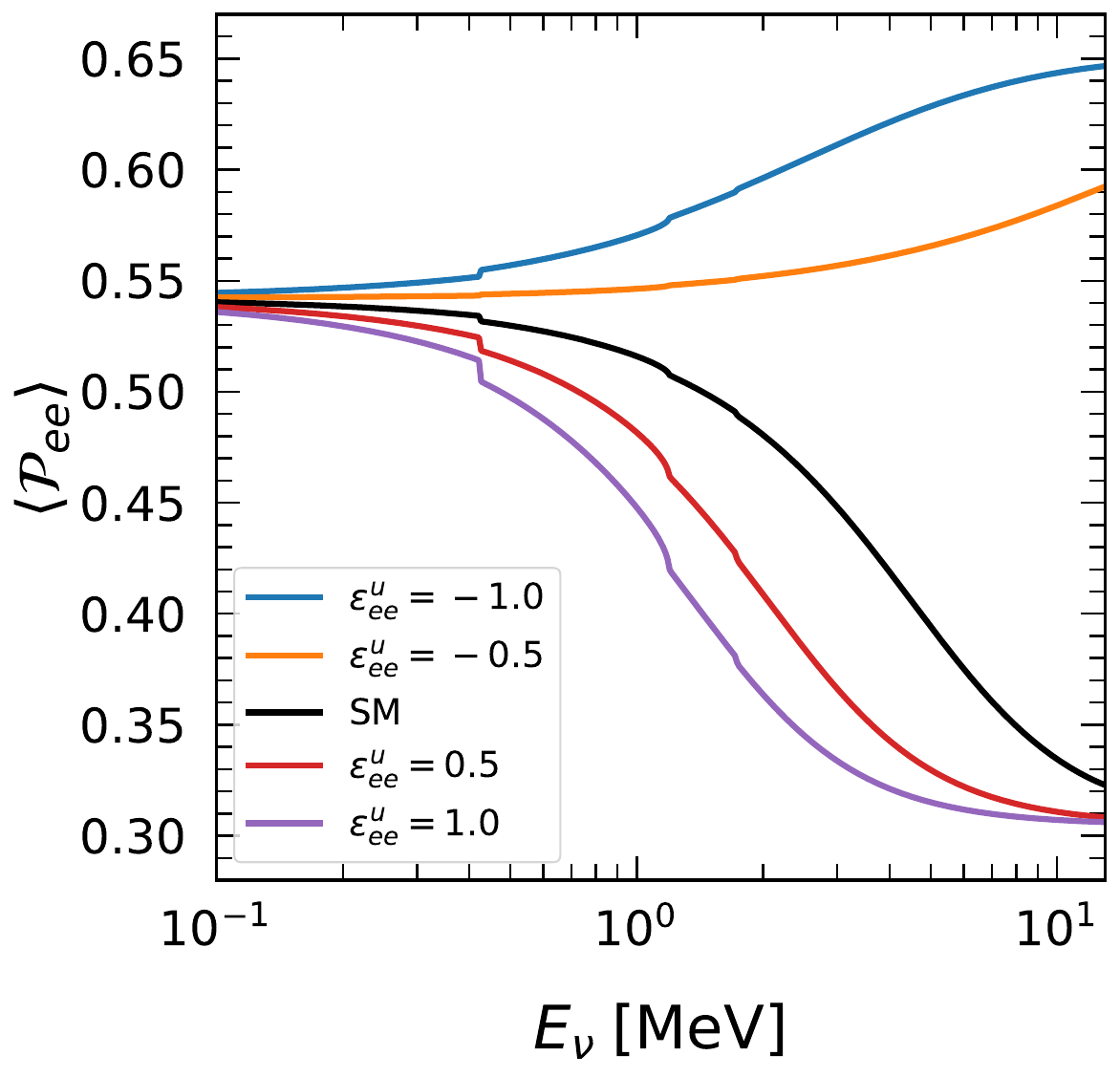}
  \caption{Averaged survival probability as a function of neutrino
    energy for the case in which only $\epsilon_{ee}^u$ has a
    non-vanishing value. This graph aims only at illustrating the
    impact of neutrino NSI on neutrino propagation in the Sun. The
    different features are related with the kinematic end-points where
    certain neutrino fluxes fade away [see
    Tab.~\ref{tab:normalization} along with
    Eq. (\ref{eq:averaged-survival-prob})].}
  \label{fig:prob_eps_ee_u}
\end{figure}
Eqs. (\ref{eq:survival-prob}) and (\ref{eq:eff-prob}) combined with
Eqs. (\ref{eq:A-B})-(\ref{eq:mixing-angle}) allow the determination of
$\mathcal{P}_{ee}(E_\nu,r)$ in terms of neutrino oscillation
parameters, electron and quark number densities and NSI
parameters. The averaged survival probability is then obtained by
integrating over $r$ taking into account the distribution of neutrino
production in the Sun \cite{Gonzalez-Garcia:2013usa}:
\begin{equation}
  \label{eq:averaged-survival-prob}
  \langle\mathcal{P}_{ee}(E_\nu)  \rangle=
  \frac{\sum_\alpha\Phi_\alpha(E_\nu)\int_0^1
    \,dr\rho(r)\,\mathcal{P}_{ee}(E_\nu,r)}
  {\sum_\alpha\Phi_\alpha(E_\nu)}\ ,
\end{equation}
where $\Phi_\alpha(E_\nu)$ refers to the $\alpha$ component of the
solar neutrino flux (with $\alpha$ running over all components) and
$\rho_\alpha(r)$ to the distribution of neutrino production. For
illustration (and only with that aim), in Fig. \ref{fig:prob_eps_ee_u}
we show an example of the averaged survival probability as a function
of the neutrino energy for the case in which all couplings but
$\epsilon_{ee}^u$ vanish. As can be seen, the new interaction can
either enhance or deplete neutrino flavor conversion depending on its
strength and on whether it reinforces or weakens the SM matter
potential. With propagation effects already discussed and summarized
in Eq. (\ref{eq:averaged-survival-prob}) we now turn to the discussion
of detection effects.
\subsection{Neutrino NSI: Detection effects}
\label{sec:detection_effects}
For consistency, the same basis used for neutrino propagation should
be used in neutrino detection as well. In doing so the effective
Lagrangian in Eq. (\ref{eq:NSI_Lag}) reads
\begin{equation}
  \label{eq:NSI_Lag}
  \mathcal{L}_\text{NSI}=-\sqrt{2}G_F\sum_{\substack{i=e,\mu,\tau\\q=u,d}}
  \overline{\widetilde\nu}_i\,\gamma_\mu P_L
  \widetilde\epsilon_{ij}^q\,\widetilde\nu_j\,\overline{q}\gamma^\mu q\ ,
\end{equation}
where
$\widetilde\epsilon^q=\mathcal{U}^T \epsilon^q\mathcal{U}\equiv
U_{13}^TU_{23}^T\epsilon^q U_{23}U_{13}$. With the couplings rotated
this way the weak-charge in the CE$\nu$NS cross section in
Eq. (\ref{eq:cross-sec-NSI}) becomes lepton flavor dependent, with the
weak-charge in initial-state flavor $i$ given by
\begin{widetext}
\begin{equation}
  \label{eq:QNSI-i}
  Q_{\nu_i}^2=
  \left[
    +
    N\left(g_V^n + \widetilde{\epsilon}_{ii}^u + 2\widetilde{\epsilon}_{ii}^d\right)
    +
    Z\left(g_V^p + 2\widetilde{\epsilon}_{ii}^u + \widetilde{\epsilon}_{ii}^d\right)
  \right]^2
  +
  \sum_{j\neq i}
  \left[
    N\left(\widetilde{\epsilon}_{ij}^u + 2\widetilde{\epsilon}_{ij}^d\right)
    +
    Z\left(2\widetilde{\epsilon}_{ij}^u + \widetilde{\epsilon}_{ij}^d\right)
  \right]^2\ .
\end{equation}
\end{widetext}
The couplings entering in the weak charge can be readily calculated
from their definition, with the rotation matrices parametrized for a
passive rotation:
$\widetilde{\epsilon}_{ij}^q=\sum_{k,\ell}\mathcal{U}_{ki}
\varepsilon^q_{k\ell}\mathcal{U}_{\ell j}$. The effects of the NSI are
then clear. By modifying the weak-charge the new interactions can
either enhance of deplete the expected reaction
rate. Eq. (\ref{eq:QNSI-i}) shows that diagonal couplings can produce
constructive or destructive interference, while off-diagonal couplings
cannot. Note that a proper definition of the flavor basis is, in
principle, not possible in the presence of flavor off-diagonal NSI
parameters. Strictly speaking then a consistent treatment of such
cases requires a density matrix formulation for the calculation of
event rates \cite{Coloma:2023ixt}. Arguably, however, differences
between the ``standard'' approach and the latter are expected to be
small provided the off-diagonal parameters are suppressed. That this
is the case is somehow expected from data, which do not sizably
deviates from the SM expectation. Thus, we adopt the standard
procedure regardless of the flavor structure of the parameters
considered.

In the two-flavor approximation, two neutrino flavors reach the
detector: $\widetilde{\nu}_e$ and $\widetilde{\nu}_\mu$. Lepton flavor
composition of the final state, however, depends on the lepton flavor
structure of the interaction. In full generality, the differential
event rate is then written as follows
\begin{equation}
  \label{eq:flavored_DRS}
  \frac{dR}{dE_r}=\sum_{k=e,\mu,\tau}\left(
    \frac{dR_{ek}}{dE_r} +
    \frac{dR_{\mu k}}{dE_r}\right)\ .
\end{equation}
Here the flavored differential event rates are obtained from
Eq. (\ref{eq:DRS}) by trading $Q_W\to Q_{\nu_i}$ and by taking into
account the survival probability, $\langle\mathcal{P}_{ee}\rangle$, in
the first term as well as the oscillation probability to the
$\widetilde{\nu}_\mu$ state, $1 - \langle\mathcal{P}_{ee}\rangle$, in
the second term. Thus, in the first (second) differential event rate
in Eq. (\ref{eq:flavored_DRS}) couplings $\widetilde{\epsilon}_{ek}^q$
($\widetilde{\epsilon}_{\mu k}^q$) contribute. These couplings are a
superposition of the NSI parameters we started with, so in a
single-parameter analysis (which we adopt in the first part in
Sec. \ref{sec:analysis}) a non-vanishing unrotated NSI parameter can
imply the presence of multiple rotated parameters at the cross section
level.
\section{Analysis and results}
\label{sec:analysis}
\begin{figure*}
  \centering
  \includegraphics[scale=0.28]{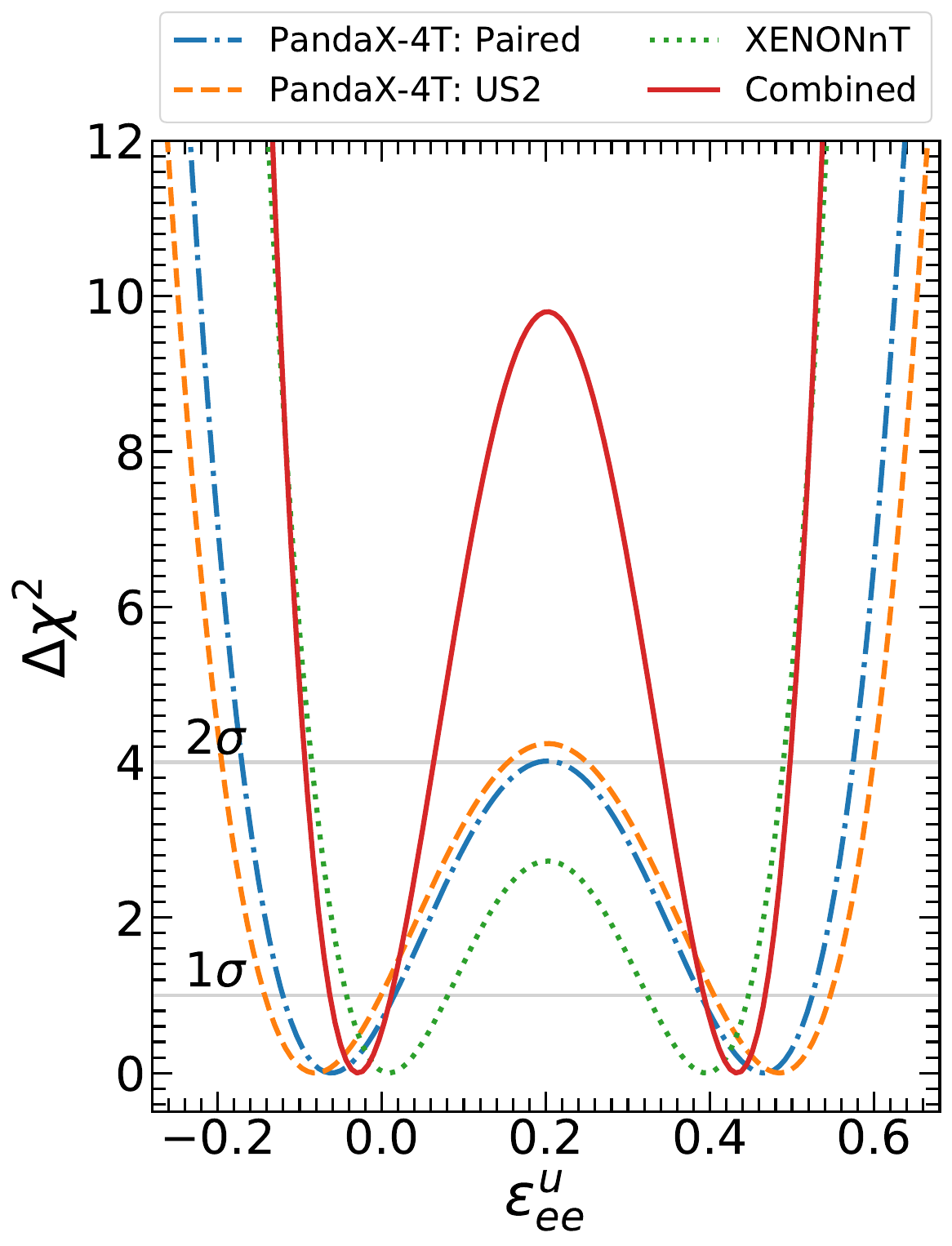}
  \includegraphics[scale=0.28]{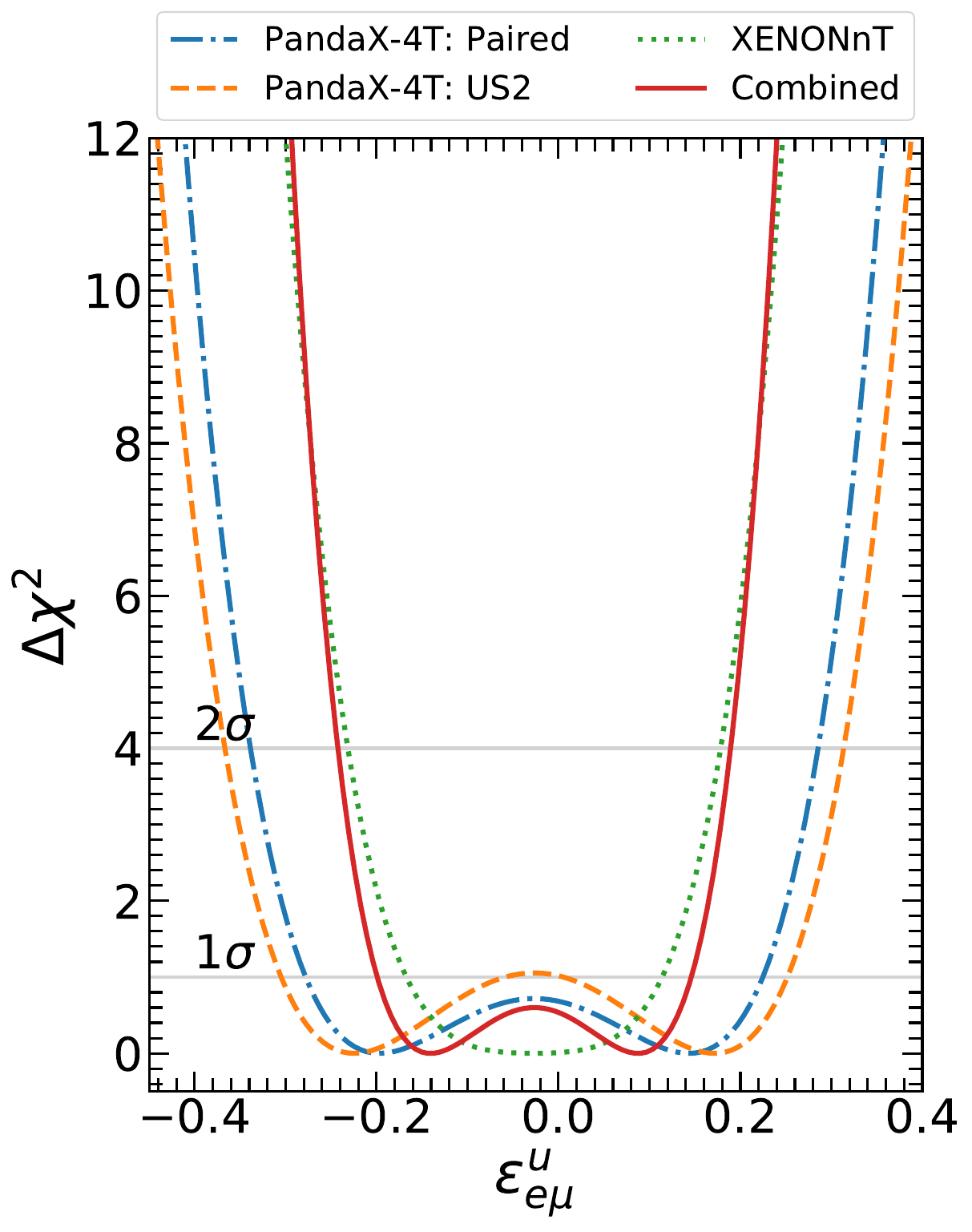}
  \includegraphics[scale=0.28]{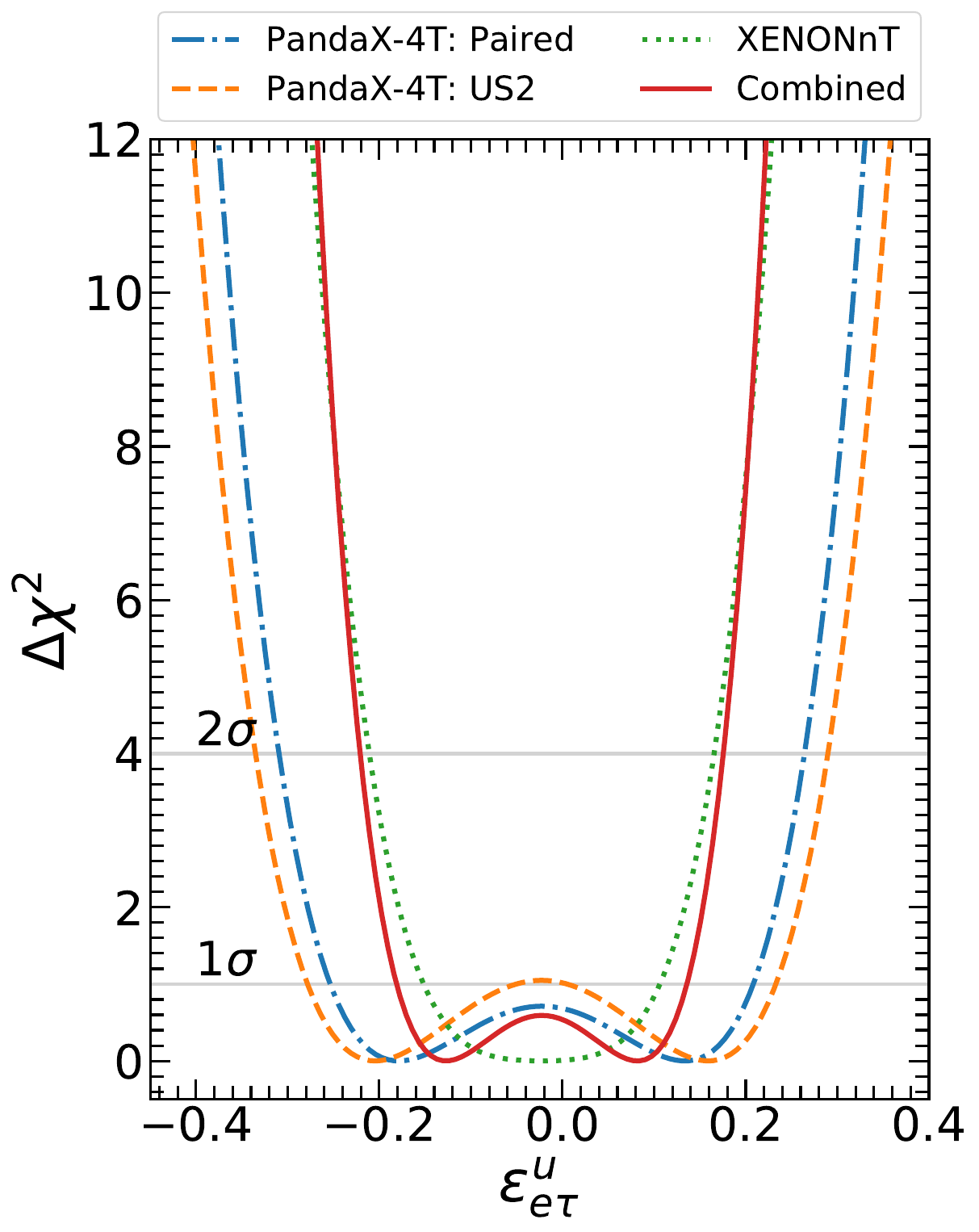}
  \includegraphics[scale=0.28]{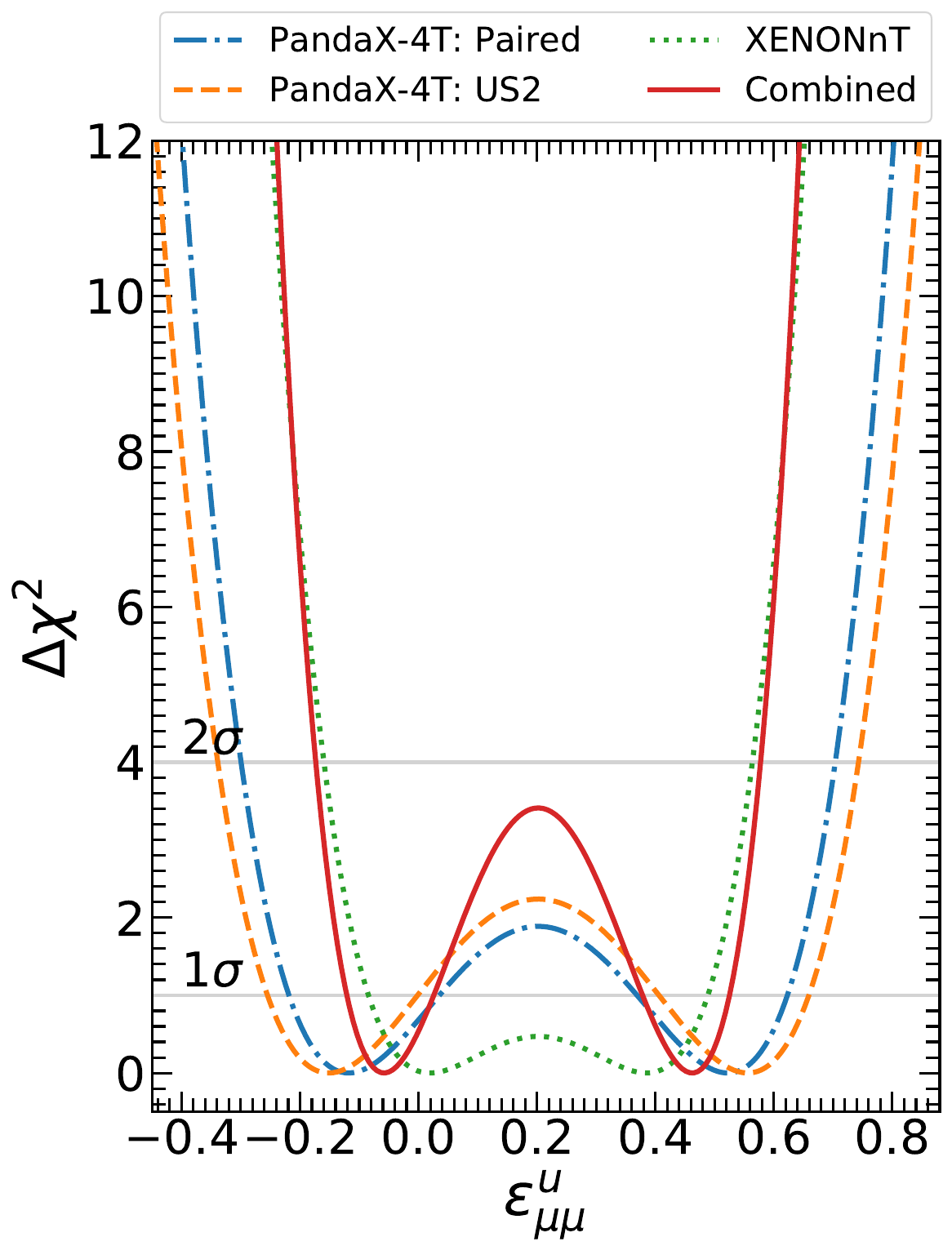}
  \includegraphics[scale=0.28]{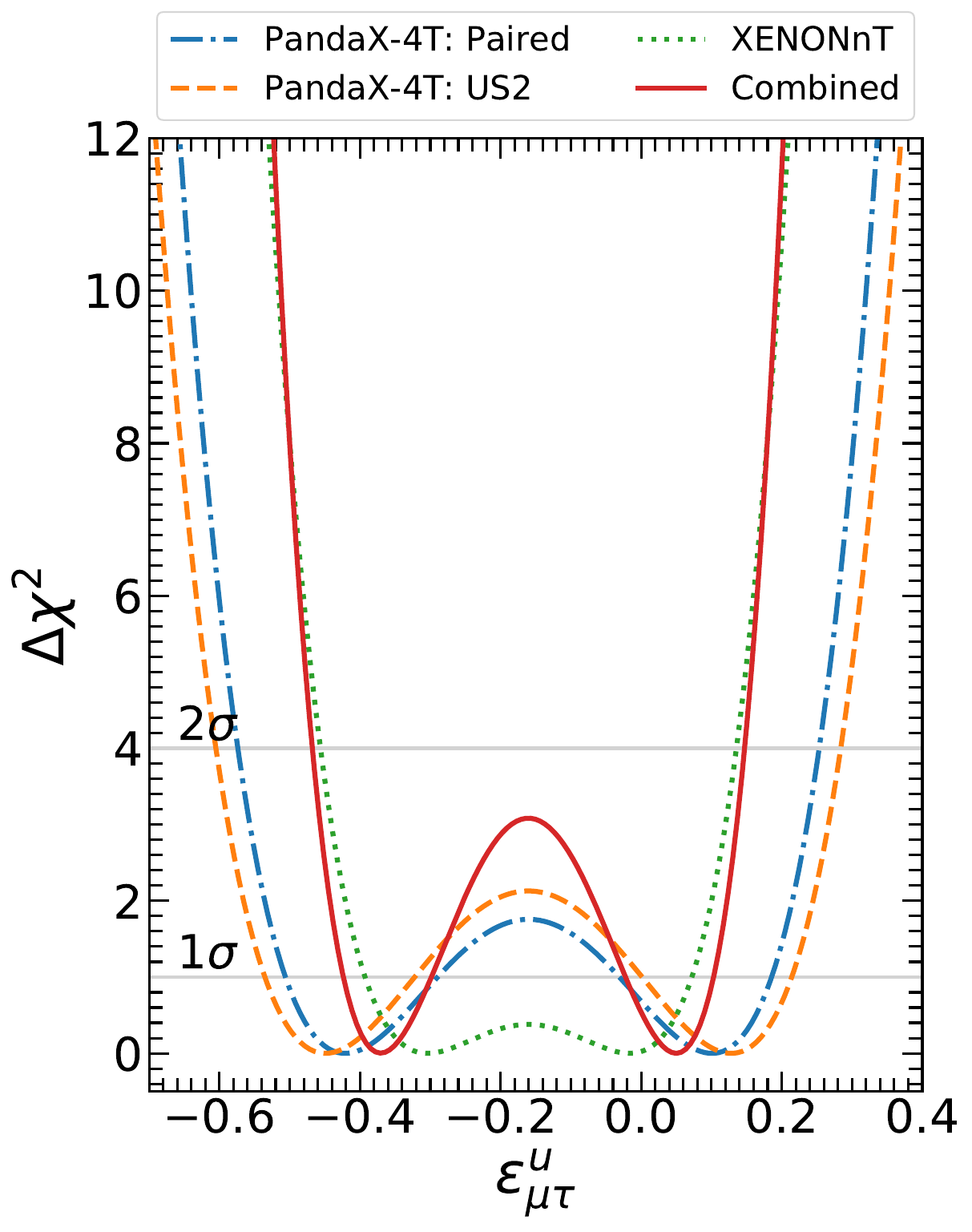}
  \includegraphics[scale=0.28]{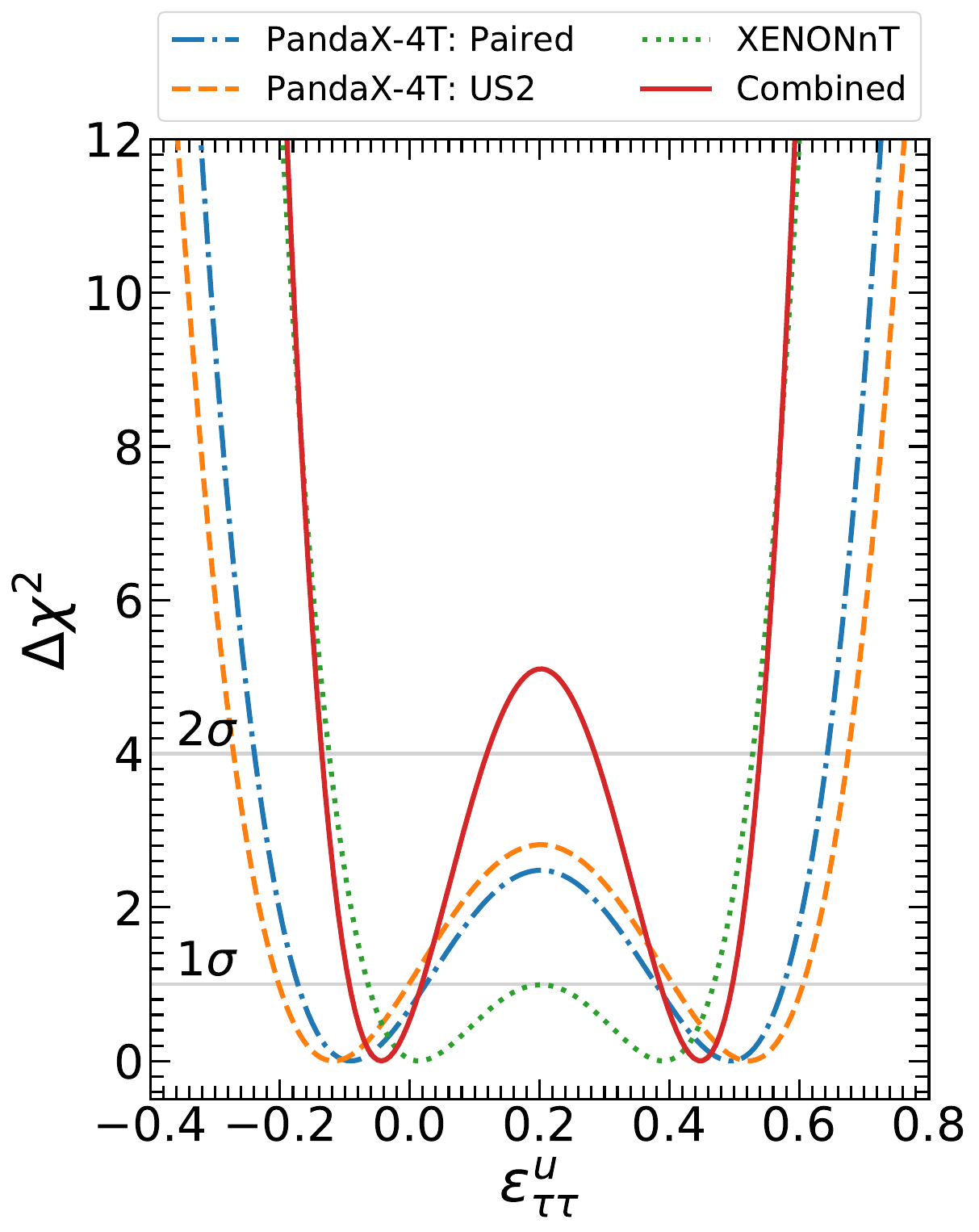}
  \caption[NSI_up_quark]{Dependence of the $\Delta\chi^2$ function on
    the up-quark NSI parameters for the PandaX-4T [paired and unpaired
    ionization-only signals (US2)] as well as for XENONnT data
    sets. Results for the combined analysis are shown as well. The
    $1\sigma$ and $2\sigma$ confidence level values (horizontal lines)
    are shown to facilitate reading.}
  \label{fig:up_quark_NSI}
\end{figure*}
The general problem of assessing the impact of neutrino NSI parameters
in neutrino-nucleus event rates involves twelve independent
couplings. It is of course a very CPU expensive problem, but not only
that. With only a few observables to rely upon, little can be said in
the most general case. For practical reasons and as well to make
contact with previous analysis, we adopt a single-parameter approach.
Towards the end of this section we consider the three lepton flavor
diagonal two-parameter cases ($\epsilon_{ee}^u$,$\epsilon_{ee}^d$),
($\epsilon_{\mu\mu}^u$,$\epsilon_{\mu\mu}^d$) and
($\epsilon_{\tau\tau}^u$,$\epsilon_{\tau\tau}^d$); as well to make
contact with what has been done previously in the literature (the $e$
and $\mu$ cases motivated by previous COHERENT data analysis). It is
worth mentioning that because of neutrino flavor mixing multi-ton DM
detectors are sensitive to $\tau$ flavor, which neither reactor nor
stopped-pion sources are. From this point of view these measurements
are unique.

We start with $u$-quark couplings and proceed by defining a simple
$\chi$-square test
\begin{equation}
  \label{eq:chi-square}
  \chi^2 = \left(\frac{R_\text{Exp} - R_\text{SM+NSI}}
    {\sigma_\text{Exp}}\right)^2\ ,
\end{equation}
where $R_\text{Exp}$ refers to PandaX-4T and XENONnT event rates
central values and (see Tab. \ref{tab:detectors_summary}) and
$R_\text{SM+NSI}$ to the SM events rates including as well NSI
contributions. Though oversimplified, such $\chi$-square statistic
allows to capture the main features of the data sets and their
sensitivity to NSI parameters. Results are shown in
Fig. \ref{fig:up_quark_NSI}. First of all, in all cases and with all
data sets two minima are found. This result follows from allowing the
NSI parameter to have positive and negative values. Because of this
range, as we have already pointed out, event rates are symmetric
around a small value. Experimental results are thus reproduced in two
non-overlapping regions of parameter space.

\begin{figure*}
  \centering
  \includegraphics[scale=0.28]{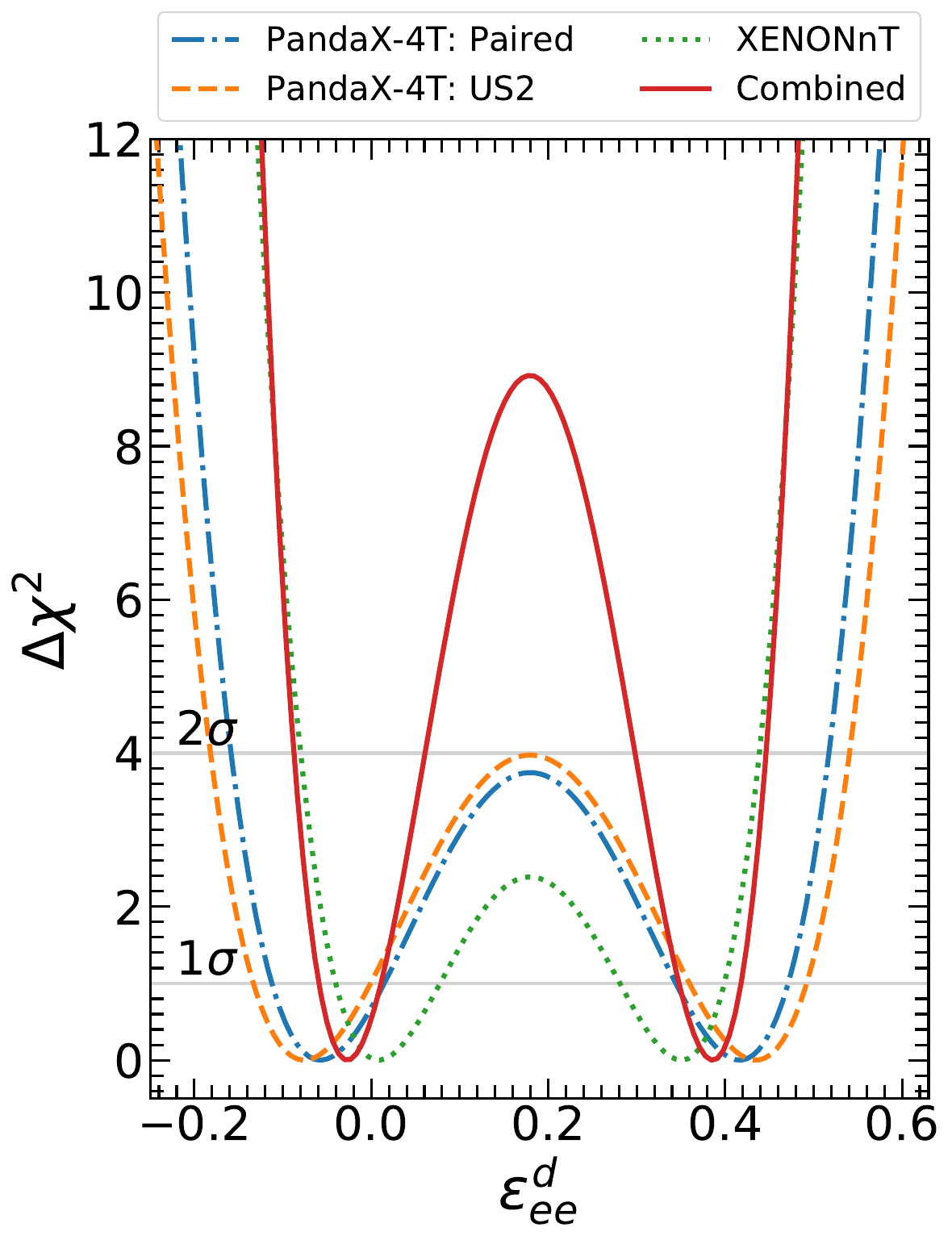}
  \includegraphics[scale=0.28]{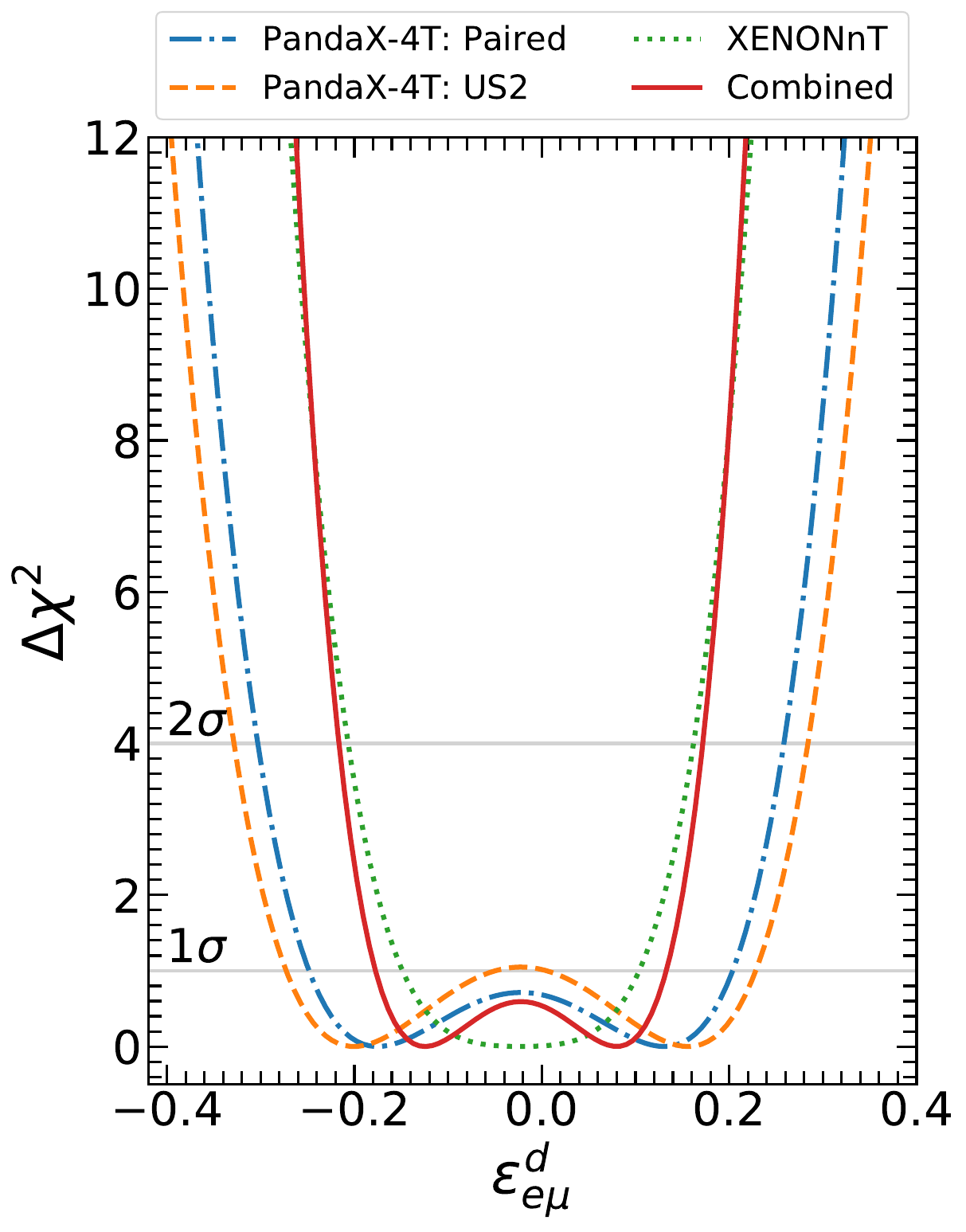}
  \includegraphics[scale=0.28]{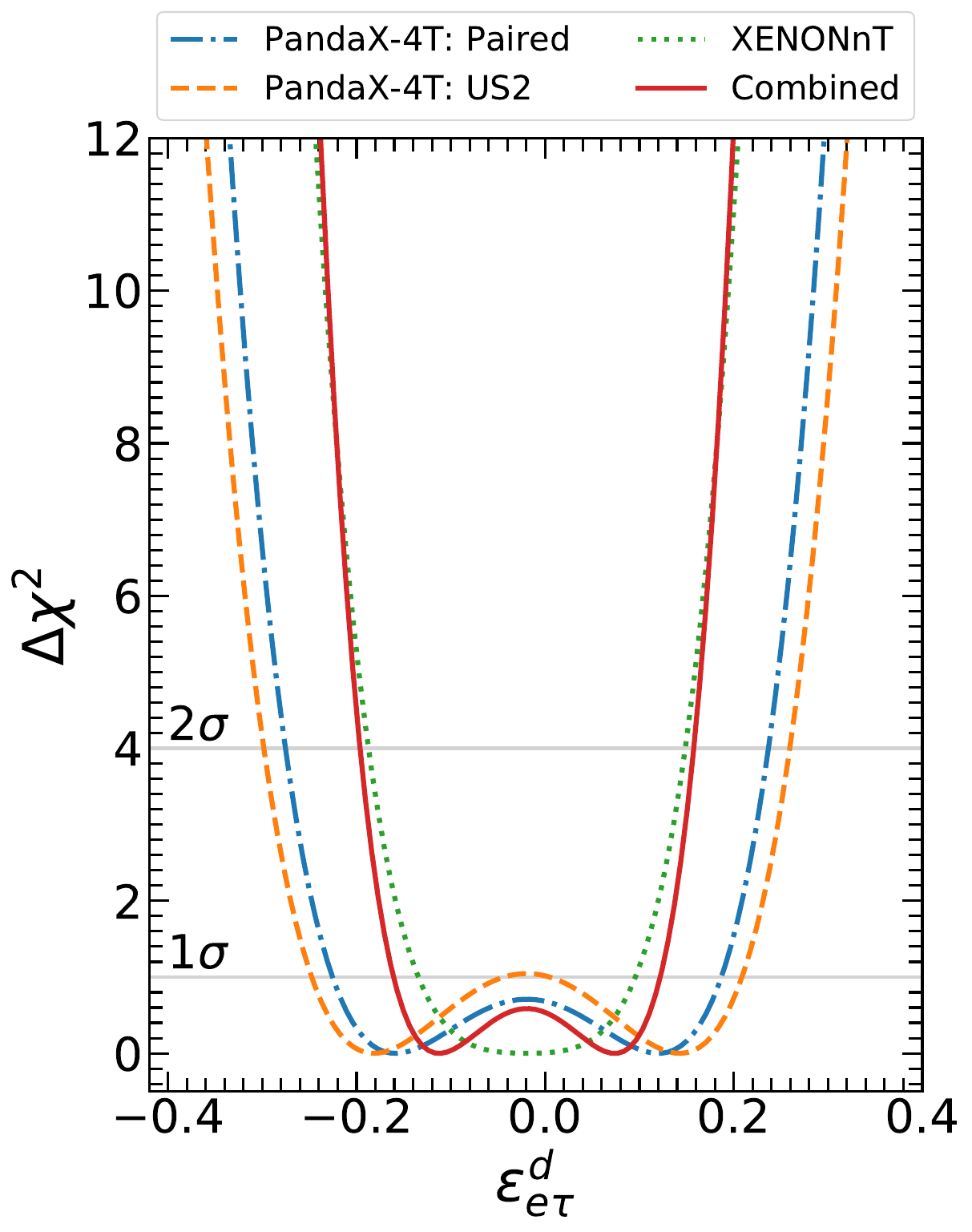}
  \includegraphics[scale=0.28]{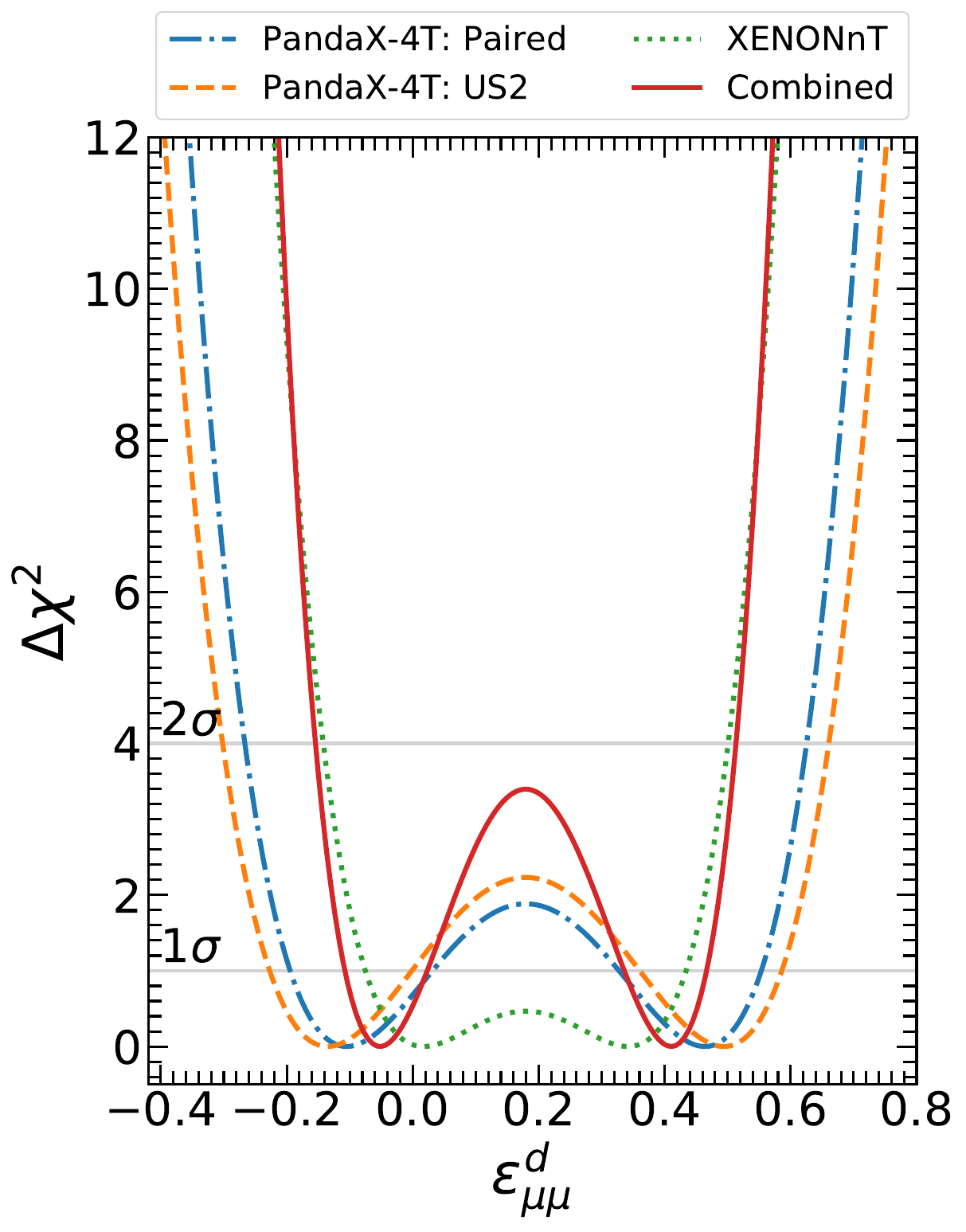}
  \includegraphics[scale=0.28]{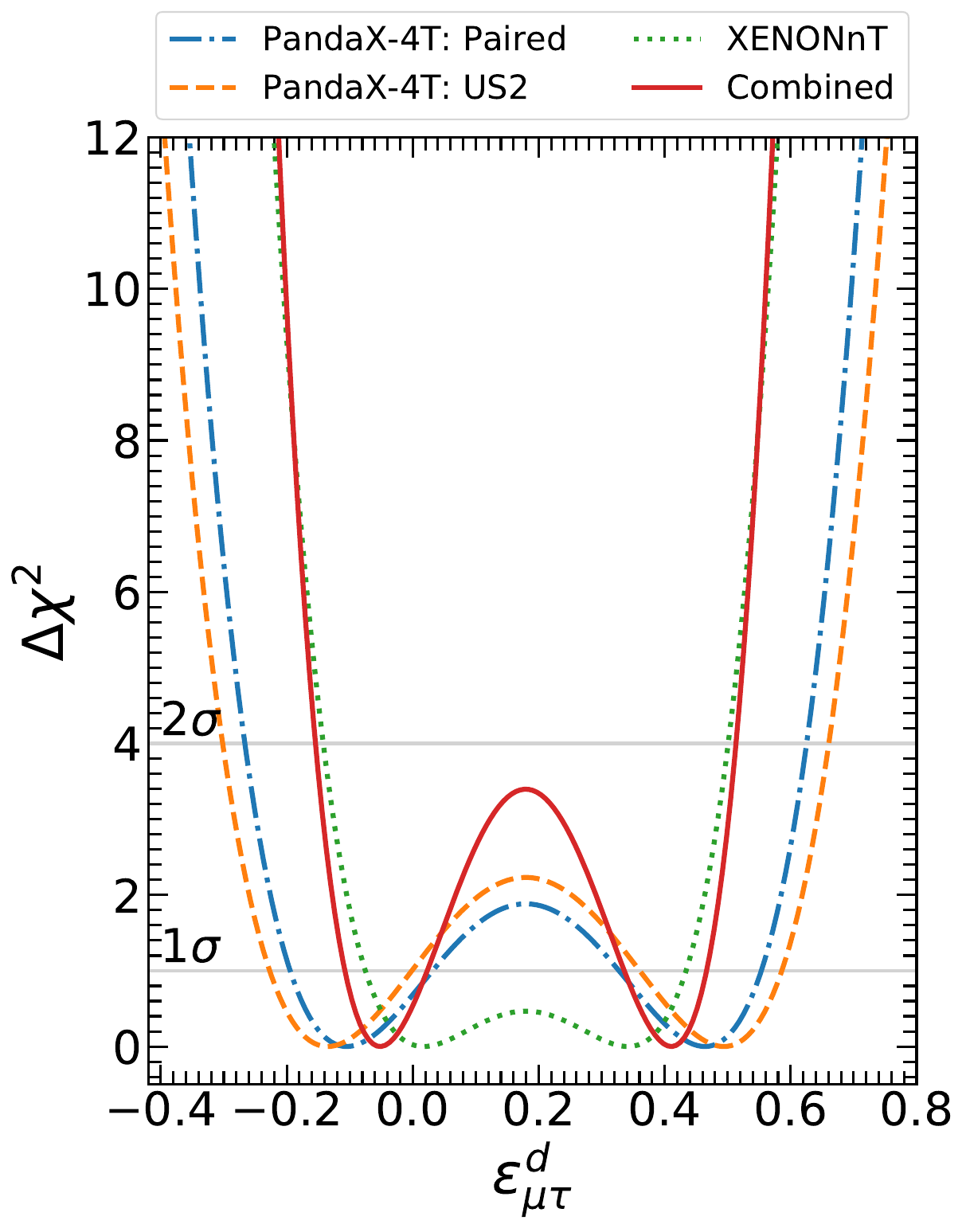}
  \includegraphics[scale=0.28]{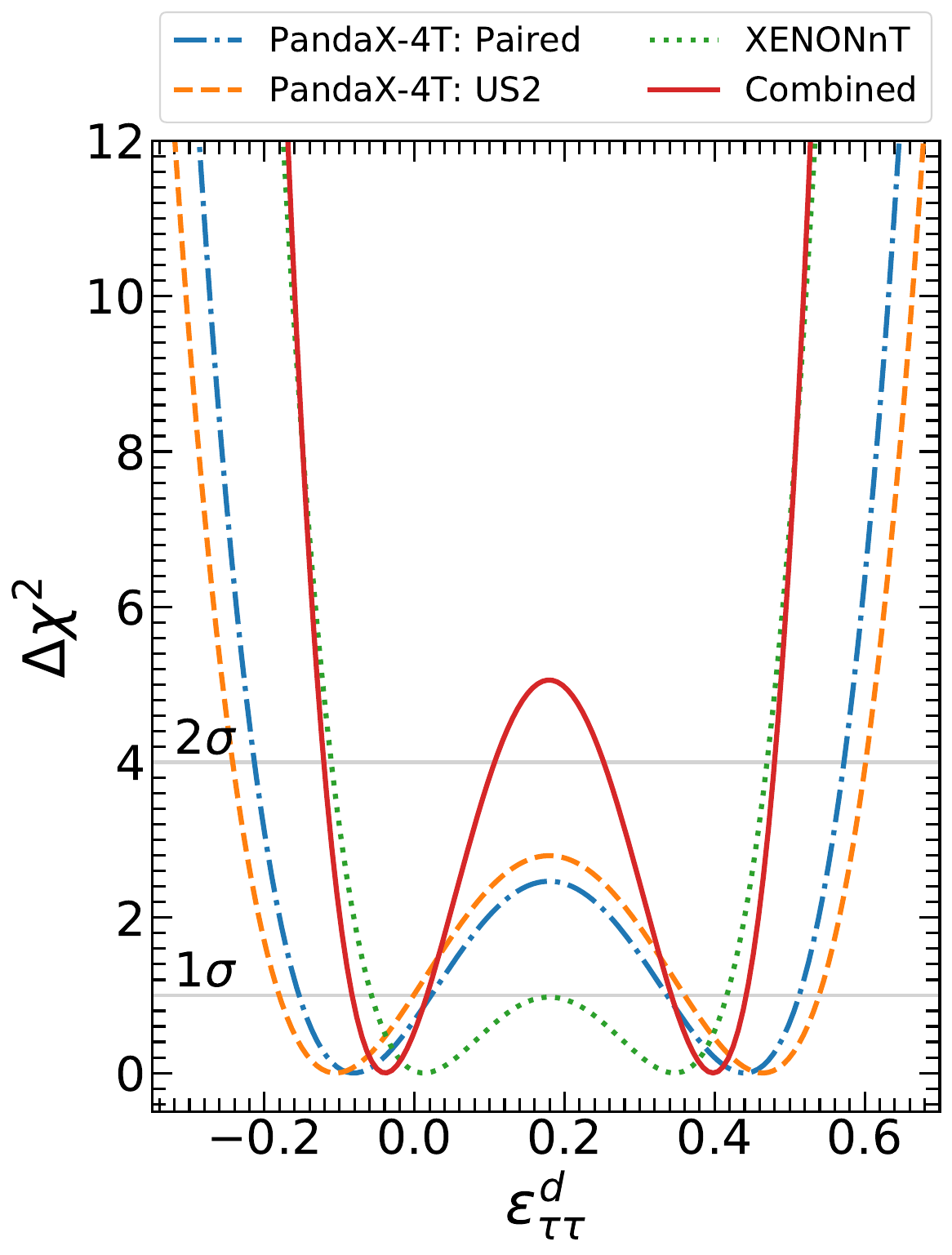}
  \caption[NSI_down_quark]{Dependence of the $\Delta\chi^2$ function
    on the down-quark NSI parameters for the PandaX-4T [paired and
    unpaired ionization-only signals (US2)] as well as for XENONnT
    data sets. Results for the combined analysis are shown as
    well. The $1\sigma$ and $2\sigma$ confidence level values
    (horizontal lines) are shown to facilitate reading.}
  \label{fig:down_quark_NSI}
\end{figure*}
One can see, however, the regions tend to be less pronounced for the
XENONnT analysis, regardless of the NSI parameter. Statistical
uncertainties are of the order of $\sim 37\%$ in all cases, so they
cannot account for this behavior. We thus understand this tendency to
be related with measured values and the SM expectation, as we now
discuss. We find for the SM predicted values 2.4:46.8:11.3 events for
paired:US2:XENONnT. Experimental ranges are on the other hand
[2.2,4.8]:[47.0,103.0]:[6.7,14.6] for paired:US2:XENONnT. So,
PandaX-4T results tend to prefer values above the SM prediction, while
the SM value expected at XENONnT is well within the measured
interval. In fact, the expected SM value is $5\%$ away from the
midrange, 10.65 events.

From the results one can see that narrower $1\sigma$ level ranges are
found for flavor-diagonal parameters. Results for flavor off-diagonal
couplings are, instead, wider. This is as well expected. At
the cross section level flavor-diagonal contributions add/subtract
linearly to the SM contribution, while flavor off-diagonal do
quadratically. Since $|\epsilon_{ij}^u|<1.0$, the diagonal components
lead to larger deviations than the off-diagonal do for larger values.

\begin{figure*}[t]
  \centering
  \includegraphics[scale=0.27]{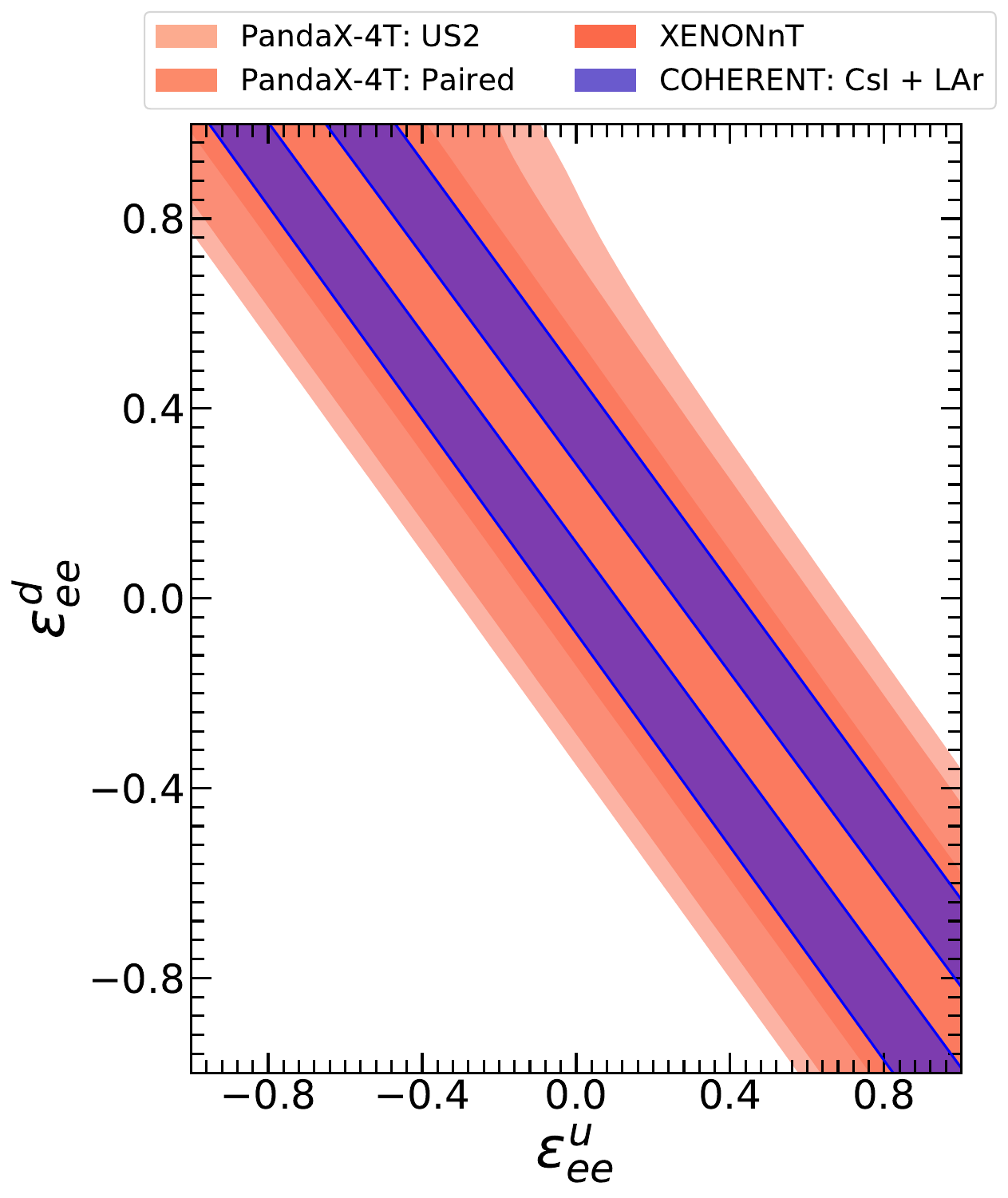}
  \includegraphics[scale=0.27]{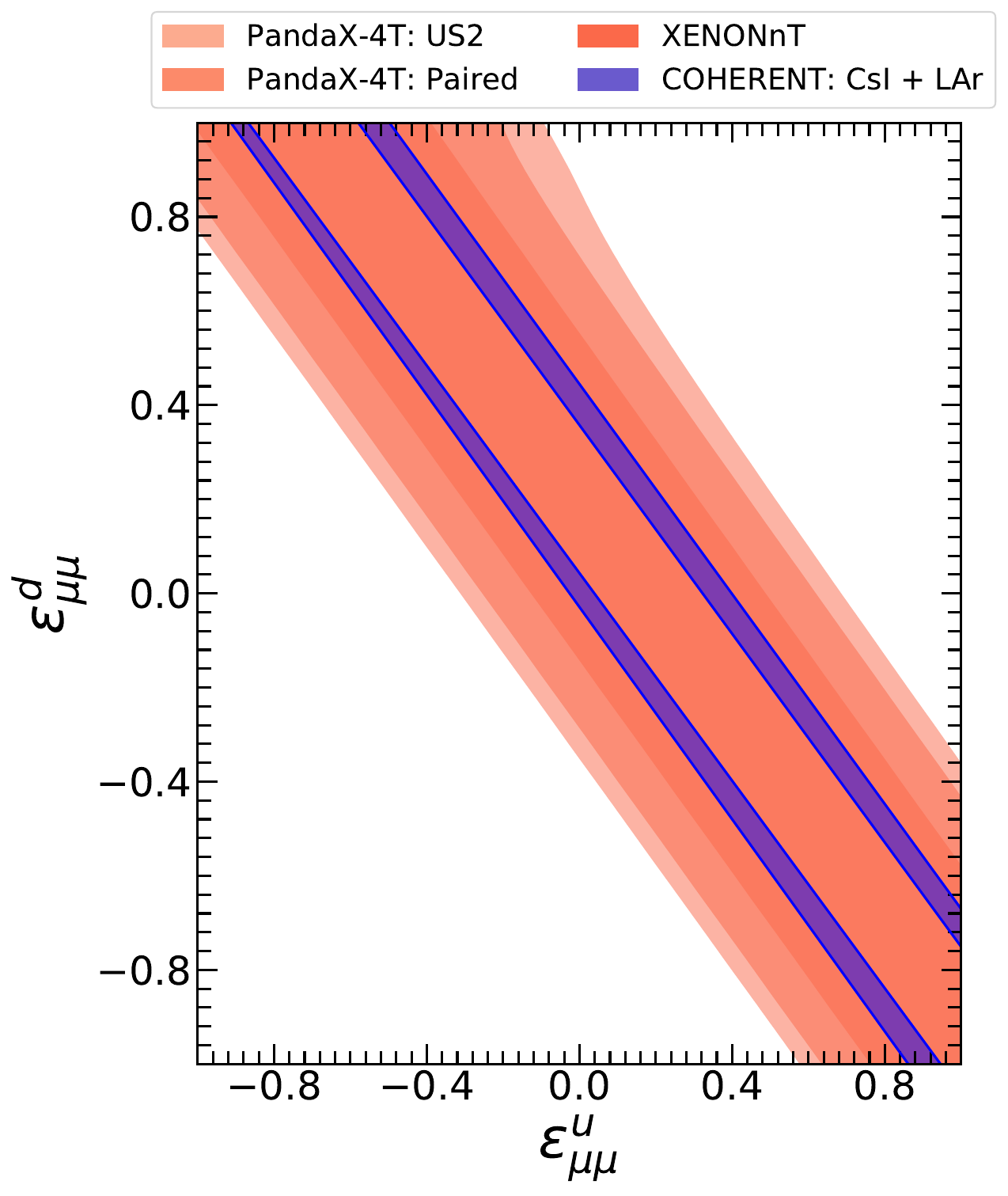}
  \includegraphics[scale=0.27]{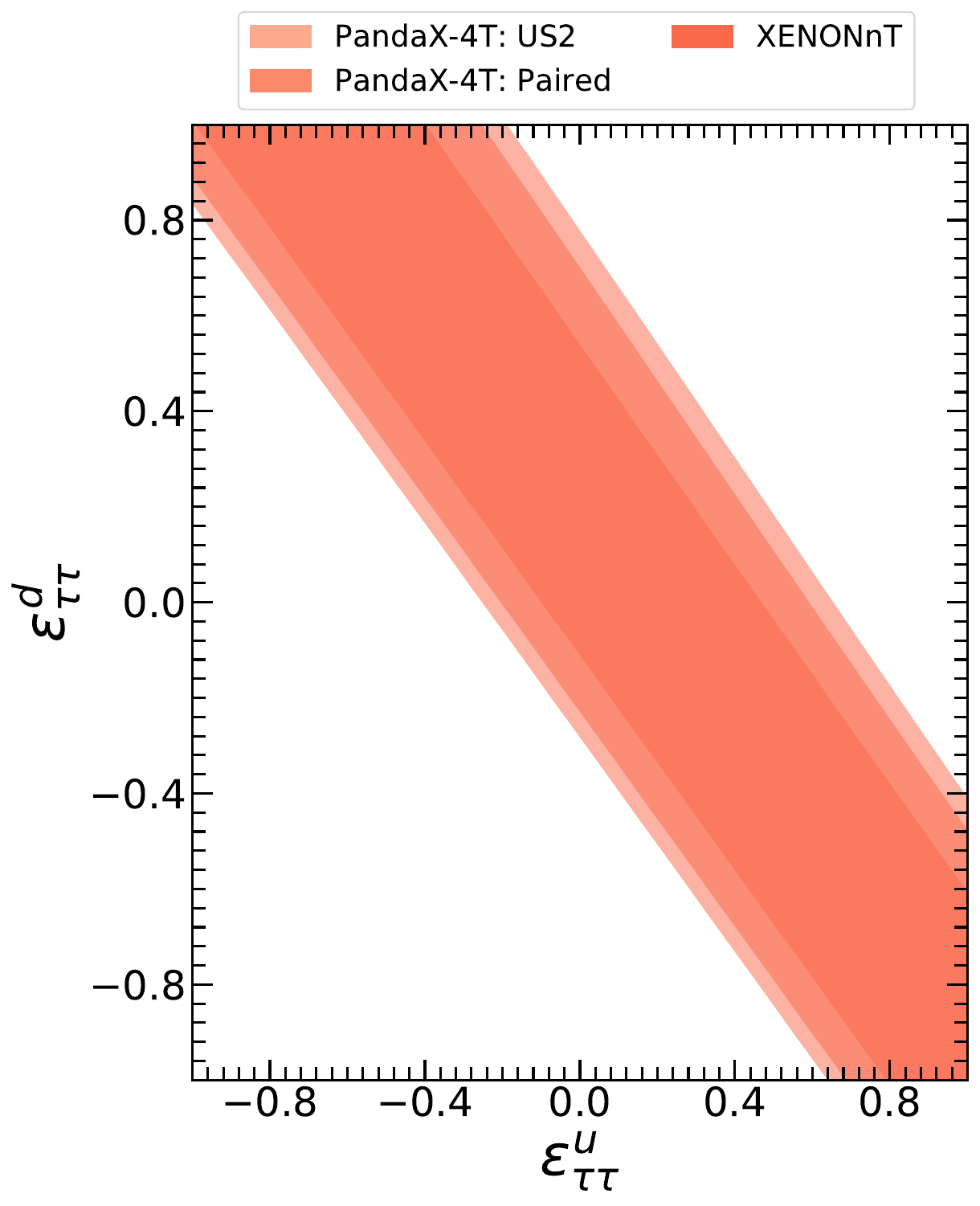}
  \caption[2_parameter_analysis]{$\Delta\chi^2$ $90\%$ CL isocontours
    in the $\epsilon_{ee}^u-\epsilon_{ee}^d$ (left graph) and
    $\epsilon_{\mu\mu}^u-\epsilon_{\mu\mu}^d$ (middle graph) and
    $\epsilon_{\tau\tau}^u-\epsilon_{\tau\tau}^d$ (left graph)
    planes. Results are shown for the PandaX-4T [paired and unpaired
    ionization-only signals (US2)] as well as for XENONnT data
    sets. For comparison results from combined analysis of COHERENT
    CsI+LAr data \cite{DeRomeri:2022twg} are shown as well. Results
    for the combined analysis have a strong overlapp with those
    from XENONnT so are not displayed. Note that COHERENT measurements
    are not sensitive to $\epsilon_{\tau\tau}^q$ NSI parameters.}
  \label{fig:two_parameter_analysis}
\end{figure*}
We provide as well results from a combined analysis, that we have
generated by constructing a combined chi-square test
$\chi^2_\text{Combined}=\chi^2_\text{Paired}+\chi^2_\text{US2}+\chi^2_\text{XENONnT}$.
These results, however, should be interpreted with certain
caution. Combining PandaX-4T and XENONnT this way is certainly
reliable, but combining paired and US2 data sets might be not because
of possible correlations. Very likely the most suitable way of combining
these data sets is through a covariance matrix. However, such an 
analysis can only be performed with the full data sets, including backgrounds. 
it can be noted that the combined analysis is dominated
by XENONnT data, with the reason being what we pointed out already:
XENONnT measurement is more inline with the SM expectation.

Results for down-quark couplings are shown in
Fig. \ref{fig:down_quark_NSI}. Differences between these results and
those found in the up-quark case are small, a result which is also
expected. From a simple inspection of Eqs. (\ref{eq:epsilonD}) and
(\ref{eq:epsilonN}) one can see that at the averaged survival
probability level they enter in the same functional form. Differences
between up and down quarks arise only through their relative abundance,
for which in the region of interest ($0.1\,R_\odot$) $Y_u$ differs by
no more than $30\%$ from $Y_d$ \cite{Vinyoles:2016djt}. At the cross
section level, the combination of down-quark couplings is different
from that from the up-quark couplings [see
Eq. (\ref{eq:QNSI-i})]. However, those differences are small and to a
certain degree smooth out at the event rate level.

We have summarized the $1\sigma$ level ranges following from these two
analyzes in Tab. \ref{tab:eps_u_ij_limits} in
App. \ref{sec:limits_summary}. It is worth comparing these results
with those derived recently from a combined analysis of COHERENT data
\cite{DeRomeri:2022twg}. For diagonal couplings these results are
rather comparable to those reported in
Ref. \cite{DeRomeri:2022twg}. More sizable deviations are found for
off-diagonal parameters, in particular for $\epsilon_{e\mu}^q$ and
$\epsilon_{\mu\tau}^q$ where the COHERENT combined analysis leads to
constraints that exceed those found here by about $20\%-50\%$. Thus,
these data sets already provide limits that are comparable with those
derived using COHERENT data. Expectations are then that with
forthcoming measurements sensitivities to possible new physics in the
neutrino sector will improve. Most relevant is the fact that contrary
to data coming from stopped-pion sources and/or reactors, measurements
from solar neutrino data are sensitive to pure $\tau$ flavor
parameters.

Finally, results for the two-parameter analysis are shown in
Fig. \ref{fig:two_parameter_analysis}. Overlaid are those derived from
COHERENT LAr+CsI combined analysis, in the two cases where they
apply. The combined analysis is not displayed because the strong
overlapp with the XENONnT data result. It is clear that COHERENT
data is moderately more sensitive to NSI effects, but results from
PandaX-4T+XENONnT already provide complementary information. We
understand this behavior as due to smaller statistical uncertainties
in the COHERENT data sets, in particular in the last CsI data set
release which largely dominates the fit \cite{DeRomeri:2022twg}.
\begin{table*}[t!]
  \centering
  \renewcommand{\arraystretch}{1.5}
  \setlength{\tabcolsep}{8pt}
  \begin{tabular}{|l||c|c|c|}\hline
    \multicolumn{4}{|c|}{\textbf{Up-type NSI couplings}}\\\hline
    {\bf Data set}&$\epsilon_{ee}^u$&$\epsilon_{e\mu}^u$&$\epsilon_{e\tau}^u$\\\hline
    Paired &$[-0.12,0.015]\oplus [0.39,0.52]$&$[-0.28,0.22]$&$[-0.25,0.21]$\\\hline
    US2 &$[-0.14,0.0011]\oplus [0.40,0.54]$&$[-0.30,-0.053]\oplus [0.0062,0.25]$&$[-0.28,-0.060]\oplus [0.0080,0.23]$\\\hline
    XENONnT &$[-0.040,0.080]\oplus [0.32,0.45]$&$[-0.17,0.11]$&$[-0.15,0.11]$\\\hline
    Combined &$[-0.060,0.010]\oplus [0.39,0.47]$&$[-0.20,0.15]$&$[-0.18,0.14]$\\\hline\hline
    {\bf Data set}&$\epsilon_{\mu\mu}^u$&$\epsilon_{\mu\tau}^u$&$\epsilon_{\tau\tau}^u$\\\hline
    Paired&$[-0.22,0.032]\oplus [0.37,0.62]$&$[-0.50,-0.29]\oplus [-0.030,0.18]$&$[-0.17,0.021]\oplus [0.38,0.57]$\\\hline
    US2&$[-0.25,-0.00053]\oplus [0.40,0.66]$&$[-0.53,-0.32]\oplus [0.00,0.21]$&$[-0.20,-0.0021]\oplus [0.40,0.60]$\\\hline
    XENONnT&$[-0.090,0.49]$&$[-0.40,0.070]$&$[-0.060,0.19]\oplus [0.22,0.47]$\\\hline
    Combined&$[-0.12,0.030]\oplus [0.38,0.53]$&$[-0.42,-0.29]\oplus [-0.020,0.11]$&$[-0.090,0.020]\oplus [0.39,0.50]$\\\hline\hline
    \multicolumn{4}{|c|}{\textbf{Down-type NSI couplings}}\\\hline
    {\bf Data set}&$\epsilon_{ee}^d$&$\epsilon_{e\mu}^d$&$\epsilon_{e\tau}^d$\\\hline
    Paired &$[-0.11,0.01]\oplus [0.34,0.47]$&$[-0.25,0.20]$&$[-0.22,0.19]$\\\hline
    US2 &$[-0.13,0.00]\oplus [0.36,0.49]$&$[-0.27,-0.05]\oplus [0.00,0.23]$&$[-0.25,-0.040]\oplus [0.010,0.21]$\\\hline
    XENONnT &$[-0.040,0.080]\oplus [0.28,0.40]$&$[-0.15,0.10]$&$[-0.13,0.090]$\\\hline
    Combined &$[-0.060,0.010]\oplus [0.35,0.42]$&$[-0.18,0.13]$&$[-0.16,0.13]$\\\hline\hline
    {\bf Data set} &$\epsilon_{\mu\mu}^d$&$\epsilon_{\mu\tau}^d$&$\epsilon_{\tau\tau}^d$\\\hline
    Paired &$[-0.20,0.030]\oplus [0.33,0.55]$&$[-0.44,-0.28]\oplus [-0.020,0.14]$&$[-0.15,0.020]\oplus [0.33,0.51]$\\\hline
    US2 &$[-0.23,0.00]\oplus [0.36,0.58]$&$[-0.46,-0.30]\oplus [0.00,0.17]$&$[-0.18,0.00]\oplus [0.36,0.54]$\\\hline
    XENONnT &$[-0.080,0.43]$&$[-0.35,0.050]$&$[-0.060,0.41]$\\\hline
    Combined &$[-0.11,0.020]\oplus [0.33,0.47]$&$[-0.37,-0.28]\oplus[-0.010,0.080]$&$[-0.080,0.020]\oplus [0.34,0.45]$\\\hline
  \end{tabular}
  \caption{$1\sigma$ CL intervals for $\epsilon_{ij}^u$ (upper table)
    and $\epsilon_{ij}^d$ (lower table) derived from PandaX-4T (paired
    and US2) and XENONnT data sets as well as from a combined analysis
    of all data. As a function of the NSI parameters, event rates tend
    to be symmetric around a value close to zero. The non-overlapping
    intervals in all cases are a result of this behavior.}
  \label{tab:eps_u_ij_limits}
\end{table*}
\section{Conclusions}
\label{sec:conclusions}
Recent measurements of nuclear recoils induced by the $^8$B solar
neutrino flux by the PandaX-4T and XENONnT collaborations have opened
a new era for both DM searches and neutrino physics. Certainly, for DM
searches this implies abandoning the free-background paradigm and
adopting new strategies in the quest for DM. For neutrino physics, on
the other hand, it provides a new landscape of opportunities that
range from precise measurements of the CE$\nu$NS cross section (at
energies below those employed in stopped-pion neutrino sources) to
searches of new physics that can potentially be hidden in the neutrino
sector. This would represent a full program, complementary to all the other CE$\nu$NS
related worldwide efforts.

With a goal of establishing sensitivity to neutrino physics, in this paper we have studied
the sensitivity of the PandaX-4T and XENONnT data sets to neutrino
NSI. We have presented a full one-parameter analysis as well as a
flavor diagonal two-parameter analysis, the latter with mainly the aim
of making contact with previous results derived using COHERENT
data.

In the one-parameter case, our findings show that with current
statistical uncertainties and exposures sensitivities to
flavor-diagonal NSI parameters are comparable to those derived using
COHERENT data. Sensitivities to flavor off-diagonal parameters are
less pronounced, but still competitive with those coming from COHERENT
measurements. In the two-parameter case, a comparison with COHERENT
recent data analysis demonstrates that with further improvements these
experiments have the potential to lead searches for new physics in the
neutrino sector through CE$\nu$NS measurements. In particular, and in
contrast to reactor or stopped-pion sources, because of neutrino
flavor mixing these experiments are sensitive to pure $\tau$ flavor
observables, providing a new channel for this flavor that is difficult to 
isolate in current solar neutrino data~\cite{Kelly:2024tvh}. 

Future data sets with improved exposures and statistical uncertainties
will improve upon the constraints we presented. For example,
increasing the exposure by a factor of 5, we checked that
sensitivities to $\epsilon_{ee}^u$ interactions may improve by about
$50\%$. Given that we are just now working with initial results from
Xenon-based DM experiments, it is likely that combined with electron
recoil measurements, data from CE$\nu$NS induced by the $^8$B solar
neutrino flux might lead searches for new physics using this type of
technology and perhaps pave the way for unexpected discoveries.

\appendix
\section{Summary of NSI parameters limits}
\label{sec:limits_summary}
In this appendix we collect the $1\sigma$ ranges for up- and
down-quark NSI parameters. Results are shown in
Tab.~\ref{tab:eps_u_ij_limits}.  For all couplings but
$\epsilon_{\tau\tau}^q$, these results should be contrasted with those
derived using COHERENT CsI+LAr data and/or Ge data
\cite{DeRomeri:2022twg,Liao:2024qoe}. This is the first time that
constraints for $\epsilon_{\tau\tau}^q$ have been derived from pure
solar neutrino CE$\nu$NS related data sets.

\section*{Acknowledgments}
The work of D.A.S. is funded by ANID under grant ``Fondecyt Regular''
1221445. L.S. and N.M. are supported by the DOE Grant No. DE-SC0010813. 
\bibliography{references}
\end{document}